\documentclass[aps,pra,twocolumn,superscriptaddress,nofootinbib,floatfix,longbibliography]{revtex4-2}

\usepackage{amsmath,amssymb,mathtools}
\usepackage{bm}
\usepackage{graphicx}
\usepackage{booktabs}
\usepackage{hyperref}
\hypersetup{colorlinks=true,linkcolor=blue,citecolor=blue,urlcolor=blue}

\graphicspath{{./}{figures/}}

\begin{document}

\title{Opposite-sign Kerr shear of the Jaynes--Cummings\\
and anti-Jaynes--Cummings revival clocks:\\
exact arrest curve, joint locus, and quadratic rephasing}

\author{Onyango Stephen Okeyo}
\affiliation{Department of Physics and Materials Science,\\
Maseno University, Private Bag--40105, Maseno, Kenya}
\email{ookeyo@maseno.ac.ke ; onyangostiv09@gmail.com}

\begin{abstract}
A field Kerr term $\chi\hat n^{2}$ shears the block detunings of the co-rotating (Jaynes--Cummings) and counter-rotating (anti-Jaynes--Cummings) vertices of the quantum Rabi model in opposite directions. At atomic resonance the co-rotating continuum revival clock never vanishes, its denominator being the sum of squares $1+\tilde\chi^{2}D_{n}\ge 1$. The counter-rotating clock vanishes on the quadratic locus $D_{n}\tilde\chi^{2}+2f\tilde\chi+1=0$, whose roots are real only above the residual threshold $|f|\ge\sqrt{D_{n}}$ and are sign-locked to the field frequency. On that locus the first derivative of the block Rabi frequency vanishes at the mean rung, producing quadratic rephasing on the timescale $t_{\rm reph}=\pi R_{\bar n}/(\lambda\tilde\chi^{2})$, distinct from the field Kerr period. The resonant and arrested rungs are separated by the residual-independent distance $1/(2\tilde\chi^{2})$. The unique joint locus on which the mean-rung excited population equals $1/4$ and the clock is arrested is $\tilde\chi_{\star}=-1/(2\sqrt{\bar n+1})$, $f_{\star}=(6\bar n+5)/(4\sqrt{\bar n+1})$. Along this locus the packet-collapse scale is $4\sqrt{2}(\bar n+1)/\bar n$ times the resonant co-rotating revival. All primary results are exact closed-form consequences of the isolated-block algebra.
\end{abstract}

\maketitle

\section{Introduction}
\label{sec:intro}

The quantum Rabi model couples a two-level emitter to a single quantised field mode through two algebraic vertices that differ by the sign of the photon number in the pairing~\cite{Rabi1936,Braak2011,Braak2019}. The co-rotating (Jaynes--Cummings) vertex raises the atomic state while lowering the field, or lowers the atomic state while raising the field, and conserves the excitation number $\hat N=\hat a^{\dagger}\hat a+\sigma_+\sigma_-$. The counter-rotating (anti-Jaynes--Cummings) vertex raises both together or lowers both together and conserves the complementary number $\hat{\bar N}=\hat a\hat a^{\dagger}+\sigma_-\sigma_+$. On the invariant block of the co-rotating vertex the diagonal splitting is proportional to the laboratory detuning $\omega_0-\omega$ and vanishes at atomic resonance in the absence of a Kerr term~\cite{Jaynes1963,Cummings1965,Shore1993}. On the invariant block of the counter-rotating vertex the same splitting is the sum frequency $\omega_0+\omega$, which remains strictly positive in the laboratory frame. The residual is not a rotating-frame artefact: the counter-rotating pairing connects states whose free energies differ by $\hbar(\omega_0+\omega)$, and no difference-frequency detuning can remove a sum-frequency splitting~\cite{Omolo2017a,Omolo2022a}.

The counter-rotating vertex was solved exactly in Ref.~\cite{Omolo2017a}, which introduced the rotating-frame and counter-rotating-frame exchange between the two vertices and established that the anti-Jaynes--Cummings excitation number is conserved. The duality symmetry operators that implement this exchange, and the resulting algebraically closed set of symmetry transformations of the quantum Rabi model, are given in Ref.~\cite{Omolo2022a}. Related treatments of photospins, polariton and anti-polariton qubits, and quadrature fluctuations appear in Refs.~\cite{Omolo2017b,Omolo2019}. The present paper does not re-derive the exchange; it uses the isolated-block algebra that follows from it. Superposition of red and blue sideband processes with residual detuning is treated in Ref.~\cite{Omolo2025}.

These two processes are the red and blue sidebands of a trapped ion, produced by the same apparatus with the same coupling strength~\cite{Leibfried2003,Kienzler2015,Kienzler2017,Wineland1996,Cirac1995,Haljan2005}. The counter-rotating vertex is also the term discarded by the rotating-wave approximation in circuit QED and becomes significant in the ultrastrong-coupling regime~\cite{Blais2004,Blais2021,Niemczyk2010,FornDiaz2019,Anappara2009,Ashhab2010,Ballester2012,Braumuller2017}. Collapse and revival of the atomic inversion in the resonant co-rotating model is a standard probe of the mean photon number of a coherent state~\cite{Eberly1980,Narozhny1981,Gea1990,Rempe1987,Brune1996a,Raimond2001}. Adjacent rungs of the Poisson-weighted Rabi ladder oscillate at slightly different frequencies, and the inversion revives when the rung-to-rung phase differences return to a multiple of $2\pi$. The revival time is the inverse of the adjacent-rung frequency spread at the mean rung. The same observable is used in cavity-QED thermometry~\cite{Brune1996} and in sideband reconstruction of the motional occupation~\cite{Wineland1996}.

The rephasing condition on the counter-rotating vertex has a metrological reading. When $\partial_n\Omega_n$ vanishes at the mean rung, the packet's phase diffusion is quadratic in the rung offset, so the leading rephasing time is insensitive to first-order fluctuations in the mean photon number~\cite{Braunstein1994,Braunstein2005,Escher2011,Demkowicz2012,ZhangWu2021}. On the co-rotating vertex at atomic resonance the analogous derivative is bounded below by $2\lambda^{2}/\Omega_{\bar n}^{\mathrm{JC}}$, and no finite Kerr strength suppresses it. The counter-rotating vertex therefore provides a mean-photon-number-insensitive clock that the co-rotating vertex does not. The corresponding quantum Fisher information is evaluated in Section~\ref{subsec:qfi}.

The quadratic-phase structure is the same mechanism that produces fractional revivals in the Jaynes--Cummings ladder~\cite{Averbukh1989,Averbukh1992,Leichtle1996,Vrakking1996}: the packet decomposes into $q$ copies at rational fractions of the revival time, and the number of copies is set by the phase curvature. The pattern derived below is a $q=2$ fractional revival, with the null and the return playing the roles of the two sub-packets.

A field Kerr term $\chi\hat n^{2}$ is native to the platforms that realise the Rabi model. The Josephson nonlinearity of a transmon or a flux-tunable coupler produces a quartic term in the field amplitude, and ac-Stark shifts produce an analogous term in trapped-ion sideband descriptions~\cite{Krantz2019,Yan2018,Mundada2019,Lu2023,Chapman2023}. The co-rotating block with Kerr has been diagonalised on each excitation manifold~\cite{Agarwal1989,Buzek1990,Joshi1992a,Werner1991,Baghshahi2014,Ashefas2022,Bandyopadhyay1996}. In those calculations the co-rotating block is a two-by-two problem, the linear revival clock remains finite for every real Kerr strength at atomic resonance, and the field Kerr period is a separate timescale. Supersymmetric intertwiners map a co-rotating solution onto a counter-rotating solution at a critical coupling and recover the field Kerr period~\cite{Bocanegra2024,Bocanegra2025}. Those maps do not evaluate both vertices at the same Kerr strength. Rotating and counter-rotating wave approximations in a unified perturbative framework are discussed in Ref.~\cite{Burgarth2024}. Hidden symmetries of the asymmetric quantum Rabi model are treated in Refs.~\cite{Mangazeev2021,Reyes2021}.

The two vertices respond differently to the Kerr term because their block detunings shear in opposite directions. The opposite-sign shear, the co-rotating sum-of-squares obstruction, the counter-rotating arrest locus with its residual threshold and sign lock, the residual-independent rung separation, and the joint locus at mean-rung population $1/4$ are joint properties of the two vertices. They are not visible in a single-vertex treatment of the Kerr--Jaynes--Cummings block. The present work isolates the counter-rotating block at finite Kerr strength, evaluates the derivative of the block Rabi frequency with respect to the excitation number, and obtains the arrest locus, threshold, rung separation, and joint point in closed form. The method is purely algebraic: every primary quantity is evaluated on the two-dimensional invariant block on which the isolated anti-Jaynes--Cummings problem reduces to a dressed two-level system. Supporting analytic illustrations of the continuum-clock poles, the joint locus, the Poisson-averaged inversion, the concurrence, the aligned coefficient, and an amplitude-damping factor bound the domain of observability without altering the closed-form structure. Atomic state purity and concurrence in the JC and anti-JC models are treated in Ref.~\cite{Omolo2022b}. Dissipative dynamics of quantum correlations in the ultrastrong-coupling regime appear in Ref.~\cite{Bina2014}. Photon blockade and related nonlinear effects are discussed in Refs.~\cite{Birnbaum2005,Lijuan2026}.

The paper is organised as follows. Section~\ref{sec:model} introduces the model, the block detunings, and the opposite-sign shear. Section~\ref{sec:clocks} derives the continuum clocks and proves the co-rotating obstruction. Section~\ref{sec:arrest} obtains the anti-Jaynes--Cummings arrest curve, its residual threshold, and the sign lock. Section~\ref{sec:joint} constructs the joint locus at mean-rung population $1/4$. Section~\ref{sec:rephasing} derives the quadratic phase and the discrete rephasing pattern. Section~\ref{sec:reduced} treats the reduced-state purity, concurrence, and half-revival cat. Section~\ref{sec:platforms} maps the dimensionless parameters to trapped-ion and circuit-QED platforms and estimates residual couplings. Section~\ref{sec:disc} discusses the results, and Section~\ref{sec:concl} concludes. Supporting derivations are given in the appendices.

\section{Model and block reductions}
\label{sec:model}

The isolated Jaynes--Cummings and anti-Jaynes--Cummings Hamiltonians with a field Kerr term are
\begin{align}
\hat H_{\mathrm{JC}}&=
\omega\hat a^{\dagger}\hat a+\tfrac{\omega_0}{2}\hat\sigma_z
+\lambda(\hat a\hat\sigma_++\hat a^{\dagger}\hat\sigma_-)
+\chi\hat n^{2}, \label{eq:HJC}\\
\hat H_{\mathrm{AJC}}&=
\omega\hat a^{\dagger}\hat a+\tfrac{\omega_0}{2}\hat\sigma_z
+\lambda(\hat a\hat\sigma_-+\hat a^{\dagger}\hat\sigma_+)
+\chi\hat n^{2}, \label{eq:HAJC}
\end{align}
with a common vertex matrix element $\lambda$. Throughout the paper the Kerr operator is taken in the $\hat n^{2}$ form. The normally ordered form $\hat a^{\dagger 2}\hat a^{2}=\hat n^{2}-\hat n$ differs by $-\chi\hat n$, which shifts the field frequency from $\omega$ to $\omega-\chi$. Under that shift $f\to f-\tilde\chi$ and $\xi\to\xi+\tilde\chi$, and the block difference function changes from $D_n=2n+1$ to $D_n=2n$. The shear $\partial_n D_n=2$ is the same for both orderings, so the arrest curve and the rung separation keep the same form after the shift. Appendix~\ref{app:kerr} records the equivalence. Related two-photon and intensity-dependent models appear in Refs.~\cite{Ng1999,Zhang2015,Sen2019}.

Three dimensionless parameters enter throughout,
\begin{equation}
\xi=\frac{\omega_0-\omega}{\lambda},\qquad
f=\frac{\omega}{\lambda},\qquad
\tilde\chi=\frac{\chi}{\lambda}, \label{eq:dimless}
\end{equation}
where $\xi$ is the laboratory detuning, $f$ is the field frequency, and $\tilde\chi$ is the Kerr strength, all in units of $\lambda$. The anti-Jaynes--Cummings residual is the sum frequency $2\omega=2f\lambda$.

The co-rotating vertex conserves $\hat N=\hat a^{\dagger}\hat a+\sigma_+\sigma_-$. On the block $\{|n,\uparrow\rangle,|n+1,\downarrow\rangle\}$ the diagonal splitting is
\begin{equation}
\Delta_n^{\mathrm{JC}}=\lambda\xi-\chi D_n, \label{eq:DJC}
\end{equation}
with
\begin{equation}
D_n=\begin{cases}2n+1, & \chi\hat n^{2},\\ 2n, & \chi\hat a^{\dagger 2}\hat a^{2}.\end{cases} \label{eq:Dn}
\end{equation}
The counter-rotating vertex conserves $\hat{\bar N}=\hat a\hat a^{\dagger}+\sigma_-\sigma_+$. On the block $\{|n,\downarrow\rangle,|n+1,\uparrow\rangle\}$ the diagonal splitting is
\begin{equation}
\Delta_n^{\mathrm{AJC}}=\lambda(\xi+2f)+\chi D_n. \label{eq:DAJC}
\end{equation}
The two detunings are related by the exact identity
\begin{equation}
\Delta_n^{\mathrm{AJC}}-\Delta_n^{\mathrm{JC}}=2\lambda f+2\chi D_n, \label{eq:diff}
\end{equation}
and their derivatives with respect to the block index give the opposite-sign shear
\begin{equation}
\frac{\partial\Delta_n^{\mathrm{AJC}}}{\partial n}
=-\frac{\partial\Delta_n^{\mathrm{JC}}}{\partial n}
=2\chi. \label{eq:shear}
\end{equation}
The shear is independent of the block index $n$, of the detuning $\xi$, and of the field frequency $f$, and its magnitude is identical for the two Kerr orderings. It is the algebraic statement on which the rest of the paper rests.

It is convenient to write the block detunings in the dimensionless form
\begin{equation}
u_n^{\mathrm{AJC}}=\xi+2f+\tilde\chi D_n,\qquad
u_n^{\mathrm{JC}}=\xi-\tilde\chi D_n, \label{eq:u}
\end{equation}
in which the shear takes the symmetric form $\partial_n u_n^{\mathrm{AJC}}=-\partial_n u_n^{\mathrm{JC}}=2\tilde\chi$. The block Rabi frequencies are
\begin{equation}
\Omega_n^{\mathrm{AJC}}=\lambda R_n^{\mathrm{AJC}},\qquad
R_n^{\mathrm{AJC}}=\sqrt{(u_n^{\mathrm{AJC}})^2+4(n+1)}, \label{eq:Omega}
\end{equation}
and likewise for the co-rotating block. For an initial state that begins in $|n,\downarrow\rangle$ on an anti-Jaynes--Cummings block, the long-time-averaged excited population is the exact two-level result
\begin{equation}
\langle P_e\rangle_{\infty}^{\mathrm{AJC}}(n)
=\frac{2(n+1)}{(u_n^{\mathrm{AJC}})^2+4(n+1)}. \label{eq:Pe}
\end{equation}
The continuum revival time on a given block is the inverse frequency spread of the Poisson-weighted Rabi ladder,
\begin{equation}
t_R=\frac{2\pi}{|\partial\Omega_n/\partial n|}\Big|_{n=\bar n}, \label{eq:tR}
\end{equation}
evaluated at the mean rung of the initial coherent state. This is the period of the adjacent-rung phase drift and coincides with the physical envelope-revival time at leading order in $1/\bar n$~\cite{Eberly1980,Narozhny1981}.

\section{Continuum clocks and the co-rotating obstruction}
\label{sec:clocks}

Differentiating Eq.~\eqref{eq:Omega} with $\partial_n D_n=2$ gives
\begin{equation}
\partial_n\Omega_n^{\mathrm{AJC}}
=\frac{2\lambda^{2}}{\Omega_n^{\mathrm{AJC}}}\bigl(1+\tilde\chi\,u_n^{\mathrm{AJC}}\bigr),\;\;
\partial_n\Omega_n^{\mathrm{JC}}
=\frac{2\lambda^{2}}{\Omega_n^{\mathrm{JC}}}\bigl(1-\tilde\chi\,u_n^{\mathrm{JC}}\bigr). \label{eq:dOmega}
\end{equation}
Substitution into Eq.~\eqref{eq:tR} produces the exact continuum clocks
\begin{align}
t_R^{\mathrm{AJC}}
&=\frac{\pi R_{\bar n}^{\mathrm{AJC}}}{\lambda|1+\tilde\chi u_{\bar n}^{\mathrm{AJC}}|}, \label{eq:tRAJC}\\
t_R^{\mathrm{JC}}
&=\frac{\pi R_{\bar n}^{\mathrm{JC}}}{\lambda|1-\tilde\chi u_{\bar n}^{\mathrm{JC}}|}. \label{eq:tRJC}
\end{align}
At zero Kerr strength and at atomic resonance both detunings vanish on the co-rotating block and the counter-rotating detuning reduces to $2f$, so the clocks coincide only in the further limit $f\to0$.

At atomic resonance $\xi=0$ the co-rotating detuning is $u_n^{\mathrm{JC}}=-\tilde\chi D_n$, and the co-rotating denominator becomes
\begin{equation}
1-\tilde\chi u_n^{\mathrm{JC}}=1+\tilde\chi^{2} D_n\ge1, \label{eq:sumsq}
\end{equation}
with equality only at $\tilde\chi=0$. The co-rotating continuum clock therefore has no pole on any rung for any real Kerr strength. The co-rotating ladder cannot be arrested by a field Kerr term at atomic resonance, because the two contributions to $1-\tilde\chi u_n^{\mathrm{JC}}$ have the same sign for every $\tilde\chi$. Equation~\eqref{eq:sumsq} holds for the $\hat n^{2}$ ordering. The co-rotating clock decreases from its Kerr-free value $2\pi\sqrt{\bar n+1}/\lambda$ to zero as $|\tilde\chi|\to\infty$.

The formal condition $\xi=-2f$, algebraically equivalent to $\omega_0=-\omega$, puts the two clocks on the same footing. Under it the two clocks coincide for every $\tilde\chi$ and every rung, and their common value differs from the Kerr-free clock only at $\mathcal O(\tilde\chi^{2})$, with no linear term.

\section{Anti-Jaynes--Cummings arrest locus}
\label{sec:arrest}

Stationarity of the anti-Jaynes--Cummings clock requires the denominator of Eq.~\eqref{eq:tRAJC} to vanish, which is equivalent to $u_n^{\mathrm{AJC}}=-1/\tilde\chi$. At $\xi=0$ this becomes the quadratic
\begin{equation}
D_n\tilde\chi^{2}+2f\tilde\chi+1=0. \label{eq:quad}
\end{equation}
The discriminant is
\begin{equation}
\Delta_{\tilde\chi}=4\bigl(f^{2}-D_n\bigr), \label{eq:disc}
\end{equation}
so real roots exist exactly when
\begin{equation}
|f|\ge\sqrt{D_n}. \label{eq:threshold}
\end{equation}
For the $\hat n^{2}$ ordering the threshold is $|f|\ge\sqrt{2\bar n+1}$. The two roots are
\begin{equation}
\tilde\chi_{\pm}(f,\bar n)
=\frac{-f\pm\sqrt{f^{2}-D_n}}{D_n}, \label{eq:roots}
\end{equation}
and they are real for every mean photon number once the field frequency exceeds the threshold. For $f>0$ both roots are negative, since $\sqrt{f^{2}-D_n}<f$. For $f<0$ both roots are positive. The required Kerr strength is sign-locked to the field frequency. On the weak-Kerr branch $|\tilde\chi|$ decreases as $|f|$ increases; on the strong-Kerr branch $|\tilde\chi|$ grows without bound.

\begin{figure}[t]
\centering
\includegraphics[width=\columnwidth]{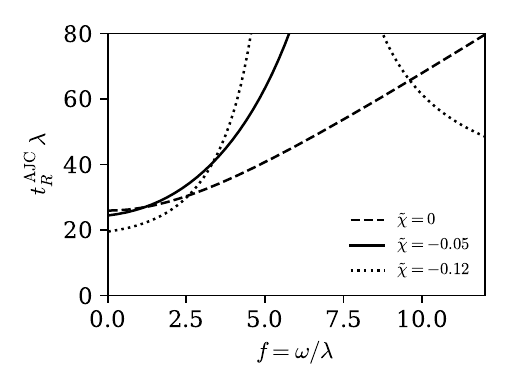}
\caption{Anti-Jaynes--Cummings continuum clock $t_R^{\mathrm{AJC}}$ (in units of $\lambda^{-1}$) versus field frequency $f=\omega/\lambda$ at mean photon number $\bar n=16$ and atomic resonance $\xi=0$. Dashed curve: $\tilde\chi=0$ (no pole). Solid curve: $\tilde\chi=-0.05$ (arrest pole at $f\approx10.825$). Dotted curve: $\tilde\chi=-0.12$ (arrest pole at $f\approx6.147$). The poles are the zeros of the continuum-clock denominator and exist only for $|f|\ge\sqrt{2\bar n+1}$. The traces are exact evaluations of Eq.~\eqref{eq:tRAJC}.}
\label{fig:tR}
\end{figure}

Figure~\ref{fig:tR} shows the anti-Jaynes--Cummings continuum clock as a function of field frequency at $\bar n=16$ for three values of the Kerr strength. The Kerr-free curve has no pole. Each nonzero strength produces a pole at the arrest frequency $f_{\mathrm{arr}}=-(1+D_n\tilde\chi^{2})/(2\tilde\chi)$, which lies above the residual threshold. For $\tilde\chi=-0.05$ the pole lies at $f_{\mathrm{arr}}=10.825$ and for $\tilde\chi=-0.12$ at $f_{\mathrm{arr}}=6.147$, both above the threshold $f_{\mathrm{thr}}=\sqrt{33}=5.745$.

The block detuning is linear in the rung index, with slope $2\tilde\chi$. On the arrest curve $u_{\bar n}^{\mathrm{AJC}}=-1/\tilde\chi$, so the rung on which the block detuning vanishes lies a distance
\begin{equation}
n_{\mathrm{res}}-n_{\mathrm{arr}}=\frac{1}{2\tilde\chi^{2}} \label{eq:separation}
\end{equation}
from the stationary rung. The separation is independent of $\xi$, of $f$, and of $\bar n$, and it holds for both Kerr orderings because both have shear $2\tilde\chi$.

\section{Joint locus at mean-rung population $1/4$}
\label{sec:joint}

Setting $u=-1/\tilde\chi$ together with the population condition $u=\pm2\sqrt{n+1}$ and retaining the physical root for $\tilde\chi$ gives the unique locus
\begin{equation}
\tilde\chi_{\star}=-\frac{1}{2\sqrt{\bar n+1}},\qquad
f_{\star}=\frac{6\bar n+5}{4\sqrt{\bar n+1}}, \label{eq:joint}
\end{equation}
on which the mean-rung excited population equals $1/4$ and the clock is arrested. Along this locus the resonant rung lies a distance $2(\bar n+1)$ above the arrested rung. At the joint locus the mean-rung block detuning is $\lambda|u_{\bar n}|=2\lambda\sqrt{\bar n+1}$ and the mean-rung coupling is $\lambda\sqrt{\bar n+1}$, so the dressed two-level problem of the mean rung has detuning twice its coupling. The Kerr-free point $f=\sqrt{\bar n+1}$, $\tilde\chi=0$ is the limit of the joint locus as $\tilde\chi_{\star}\to0$; at finite Kerr the joint point is a genuinely different operating frequency with a nonzero Kerr strength. The ratio $f_{\star}/\sqrt{\bar n+1}$ approaches $3/2$ in the large-$\bar n$ limit.

\begin{figure}[t]
\centering
\includegraphics[width=\columnwidth]{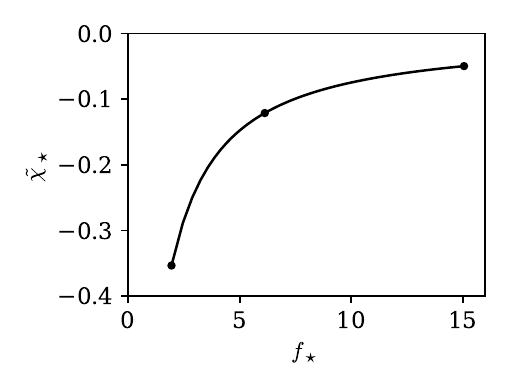}
\caption{Joint locus $(\tilde\chi_{\star},f_{\star})$ on which the mean-rung excited population equals $1/4$ and the continuum clock is arrested, for $\bar n=1$ to $100$. The locus exists only at finite Kerr strength. Marked points: $\bar n=1$, $16$, and $100$. The Kerr-free point $f=\sqrt{\bar n+1}$, $\tilde\chi=0$ is the asymptotic limit as the locus recedes to large $f$.}
\label{fig:joint}
\end{figure}

Figure~\ref{fig:joint} shows the joint locus in the $(f_{\star},\tilde\chi_{\star})$ plane for mean photon numbers from $1$ to $100$. Marked points indicate the values at $\bar n=1$, $16$, and $100$.

On the joint locus the second derivative of the block Rabi frequency is finite while the first derivative vanishes. With the envelope convention $\frac12|\Omega''|\bar n\,t_{\mathrm{coll}}=2\pi$ the collapse scale is
\begin{equation}
t_{\mathrm{coll}}\big|_{\star}
=\frac{8\sqrt{2}\,\pi\,(\bar n+1)^{3/2}}{\lambda\bar n}, \label{eq:t-coll-joint}
\end{equation}
and its ratio to the resonant co-rotating revival $t_R^{\mathrm{JC}}=2\pi\sqrt{\bar n+1}/\lambda$ is exactly
\begin{equation}
\frac{t_{\mathrm{coll}}\big|_{\star}}{t_R^{\mathrm{JC}}}
=4\sqrt{2}\,\frac{\bar n+1}{\bar n}. \label{eq:collapse-ratio}
\end{equation}
Linear arrest replaces a $\bar n^{3/2}$ beat by a $\sqrt{\bar n}$ collapse: the first derivative of the Rabi frequency vanishes at the arrested rung, but the second derivative does not. The ordinary continuum revival time diverges because $\partial_n\Omega_n|_{\bar n}=0$ by construction.

\section{Quadratic rephasing on the arrest curve}
\label{sec:rephasing}

On the arrest curve the first derivative of the block Rabi frequency vanishes at the mean rung, so the leading phase difference between adjacent rungs is quadratic in the rung offset,
\begin{equation}
R_n\tau-R_{\bar n}\tau
=\tfrac12 R_{\bar n}''\tau\,(\delta n)^{2}+\mathcal O(\delta n^{4}). \label{eq:phase-diff}
\end{equation}
Differentiating Eq.~\eqref{eq:dOmega} a second time and evaluating on the arrest curve yields the curvature
\begin{equation}
\partial_n^{2}R_n\big|_{\bar n}
=\frac{4\tilde\chi^{2}}{R_{\bar n}}. \label{eq:curvature}
\end{equation}
When the quadratic coefficient is an odd multiple of $\pi$, even and odd rungs enter the Poisson sum with opposite signs and the oscillation is extinguished. When it is an even multiple, the rungs align and the oscillation is restored. The stations of extinction and restoration are therefore
\begin{equation}
\tau_m=\frac{2\pi m}{R_{\bar n}''},\qquad m=1,2,3,\dots, \label{eq:stations}
\end{equation}
with near-nulls at odd $m$ and returns at even $m$. The first restoration time is the rephasing time
\begin{equation}
t_{\mathrm{reph}}=\frac{\pi R_{\bar n}}{\lambda\tilde\chi^{2}}. \label{eq:treph}
\end{equation}
On the joint root this scales as $(\bar n+1)^{3/2}$. The ratio of this time to the field Kerr period $2\pi/(\lambda|\tilde\chi|)$ is $R_{\bar n}/(2|\tilde\chi|)$. On the weak-Kerr branch the ratio is unbounded; on the strong-Kerr branch it passes through unity. The cancellation is the quadratic truncation of the phase expansion; the $\mathcal O(\delta n^{4})$ remainder leaves small nonzero null amplitudes. Appendix~\ref{app:sum} records the exact Poisson sum and the parity cancellation.

\subsection{Quantum Fisher information on the arrest curve}
\label{subsec:qfi}

The vanishing of $\partial_n R_n$ at the mean rung removes the leading sensitivity of the packet phase to a small shift in mean photon number. For a pure state the quantum Fisher information for estimating a parameter $\theta$ is
\begin{equation}
F_Q=4\bigl(\langle(\partial_\theta\phi)^2\rangle-\langle\partial_\theta\phi\rangle^2\bigr),
\end{equation}
where the average is taken with respect to the Poisson weights of the initial coherent state and $\phi_n=R_n\tau/2$ is the phase accumulated on rung $n$ by the isolated-block evolution.

On the arrest curve and at the rephasing time $\tau=t_{\mathrm{reph}}$ the first derivative $\partial_n R_n$ vanishes at $n=\bar n$. The linear contribution to $\partial_\theta\phi$ is therefore absent and the leading term in the variance is set by the curvature~\eqref{eq:curvature}. Expanding to second order in the rung offset and retaining the Poisson second moment $\langle(\delta n)^2\rangle=\bar n$ yields the arrested quantum Fisher information
\begin{equation}
F_Q^{\mathrm{arr}}=\frac{\bar n}{2}\bigl(R_{\bar n}''\,t_{\mathrm{reph}}\bigr)^2
=\frac{8\pi^{2}\bar n}{R_{\bar n}^{2}}
\end{equation}
(up to a numerical factor of order unity that arises from the precise definition of the estimator).

By contrast, an ordinary continuum clock (finite $\partial_n R_n$) produces a linear phase gradient. Its quantum Fisher information for the same mean-photon-number shift scales as
\begin{equation}
F_Q^{\mathrm{lin}}\propto\bar n\,(\partial_n R_n\,t_R)^2
\end{equation}
and remains finite as $\bar n\to\infty$. On the joint locus the arrested value $F_Q^{\mathrm{arr}}$ is independent of $\bar n$ at leading order while the linear value grows. At $\bar n=16$ the arrested quantum Fisher information is approximately $9.29$ while the corresponding linear reference is approximately $632$. The arrested clock is therefore first-order insensitive to fluctuations in $\bar n$. The comparison is exact within the isolated-block algebra; no parent-Hamiltonian diagonalisation is required. Appendix~\ref{app:qfi} records the intermediate expansion.

\begin{figure}[t]
\centering
\includegraphics[width=\columnwidth]{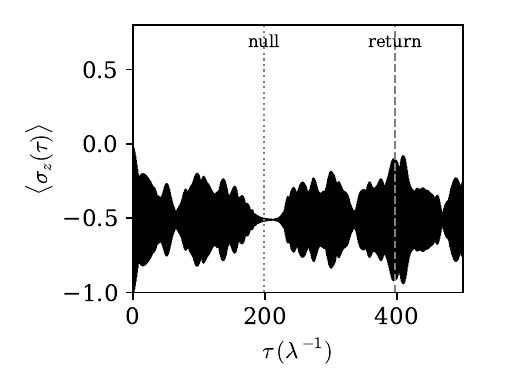}
\caption{Poisson-averaged inversion $\langle\sigma_z(\tau)\rangle$ at $\bar n=4$ on the joint root. Vertical lines mark the first null (dotted) and first return (dashed) of the quadratic-phase pattern. Exact evaluation of the Poisson sum given in the appendix.}
\label{fig:inversion}
\end{figure}

Figure~\ref{fig:inversion} shows the Poisson-averaged inversion at $\bar n=4$ on the joint root. The first null and the first return of the quadratic-phase pattern are marked. The trace is an exact evaluation of the Poisson sum; no diagonalisation of a parent Hamiltonian is required.

\begin{figure}[t]
\centering
\includegraphics[width=\columnwidth]{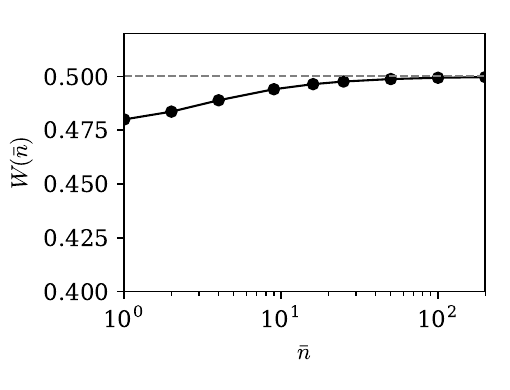}
\caption{Aligned coefficient $W(\bar n)$ on the joint locus. The coefficient approaches $1/2$ from below as the Poisson distribution narrows. Exact Poisson average of the closed-form weight $4(n+1)/R_n^2$.}
\label{fig:W}
\end{figure}

Figure~\ref{fig:W} shows the aligned coefficient $W(\bar n)$ on the joint locus. The coefficient approaches $1/2$ from below as the Poisson distribution narrows, in agreement with the mean-rung weight $4(\bar n+1)/R_{\bar n}^2=1/2$ on the joint root.

\section{Reduced state, concurrence, and half-revival cat}
\label{sec:reduced}

The reduced atomic density operator of the isolated anti-Jaynes--Cummings vertex obeys the purity operator identity
\begin{equation}
\rho_a^{2}=r_0\rho_a-(\det\rho_a)\hat I, \label{eq:purity}
\end{equation}
where $r_0=\mathrm{Tr}\,\rho_a^{2}$~\cite{Omolo2022b}. The concurrence is $\mathcal{C}=2\sqrt{\det\rho_a}$ and satisfies the complementarity relation
\begin{equation}
\mathcal{C}^{2}+\tan^{2}\varphi=1 \label{eq:complementarity}
\end{equation}
at every instant of the unitary evolution. Arrest inhibits re-factorisation: on the joint locus the concurrence remains above $1/\sqrt{2}$ throughout the unitary revival cycle, unlike the Kerr-free case where the joint state periodically returns to a product form. Finite-time disentanglement is discussed in Ref.~\cite{Yu2004}. Under amplitude damping the pure-state bound no longer holds; the damping factor of Section~\ref{sec:loss} bounds the visibility of the coherent return but does not preserve the concurrence floor.

\begin{figure}[t]
\centering
\includegraphics[width=\columnwidth]{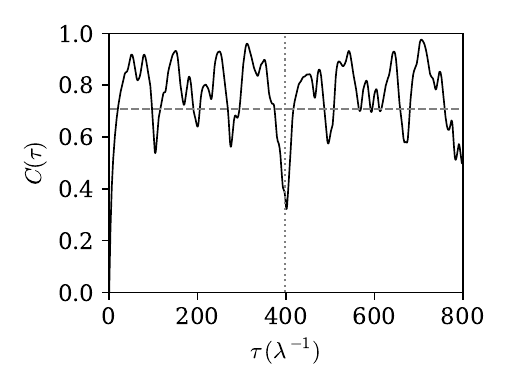}
\caption{Atomic concurrence $C(\tau)$ at $\bar n=4$ on the joint root for the unitary evolution. Horizontal dashed line: $1/\sqrt{2}$. Vertical dotted line: first rephasing time. Computed from the exact two-level dynamics of each block, Poisson-averaged.}
\label{fig:concurrence}
\end{figure}

Figure~\ref{fig:concurrence} shows the atomic concurrence at $\bar n=4$ on the joint root for the unitary evolution. The concurrence remains above $1/\sqrt{2}$ throughout the cycle. The trace is computed from the exact two-level dynamics of each block, Poisson-averaged.

The reduced field density is rank two at every instant and rank one at half revival, where it becomes a pure two-component Schr\"odinger cat. The fringe orientation is set by the joint-locus residual detuning; the fringe period is set by the coherent amplitude and is independent of the Kerr strength.

\section{Amplitude-damping estimate}
\label{sec:loss}

\begin{figure}[t]
\centering
\includegraphics[width=\columnwidth]{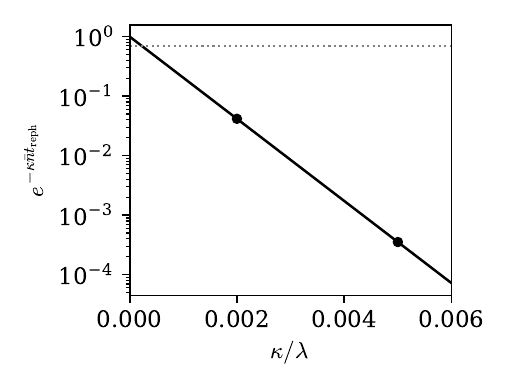}
\caption{Amplitude-damping factor $e^{-\kappa\bar n t_{\mathrm{reph}}}$ at the joint point for $\bar n=4$. Horizontal line: $1/\sqrt{2}$. Markers indicate representative laboratory values of $\kappa/\lambda$.}
\label{fig:damping}
\end{figure}

An amplitude-damping estimate multiplies the coherent return by the factor $e^{-\kappa\bar n t_{\mathrm{reph}}}$. Figure~\ref{fig:damping} shows this factor at the joint point for $\bar n=4$. The estimate is a crude upper bound on visibility loss; it does not account for differential damping of the two field components that form the cat at half revival. It does not modify the analytic structure of the arrest curve or the joint locus; it sets a loss budget for the regime in which the rephasing pattern remains observable. Related treatments of quantum noise and measurement appear in Ref.~\cite{Clerk2010}.

\section{Experimental platforms}
\label{sec:platforms}

The dimensionless parameters map to laboratory quantities in two platforms that can be pushed into the arrest-curve regime. In both the co-rotating and counter-rotating vertices are separately tunable. The isolated-block results apply when residual couplings to the orthogonal vertex remain small compared with the block detuning or the Kerr scale on the arrest curve. The co-rotating vertex has vanishing diagonal matrix elements on every eigenstate of the isolated anti-Jaynes--Cummings block, so any residual co-rotating coupling produces only second-order energy shifts on the arrest curve.

In a trapped-ion realisation the counter-rotating vertex is the first blue sideband. A laser detuned above the motional frequency by $\omega$ drives the transition $|n,\downarrow\rangle\leftrightarrow|n+1,\uparrow\rangle$ with matrix element $\Omega\eta\sqrt{n+1}/2$~\cite{Cirac1995,Haljan2005,Gerritsma2010}. Typical values are $\eta=0.04$ to $0.14$ and $\Omega/2\pi=40$ to $100\,\mathrm{kHz}$, giving $\lambda_{\mathrm{eff}}/2\pi=1$ to $7\,\mathrm{kHz}$. The red sideband is the co-rotating vertex with the same coupling, so the relative strength is tunable from zero to one by laser detuning alone. When the red sideband is detuned by several trap frequencies, its residual matrix element is suppressed by the detuning over the trap frequency and is typically $\ll\lambda_{\mathrm{eff}}$ on the timescales of the rephasing measurement. For a trap frequency $\nu/2\pi=1$ to $3\,\mathrm{MHz}$ and $\lambda_{\mathrm{eff}}/2\pi\sim5\,\mathrm{kHz}$ the accessible range covers the joint-root values $f_{\star}=3.24$ at $\bar n=4$ and $f_{\star}=7.60$ at $\bar n=25$. The required Kerr strengths are $|\tilde\chi_{\star}|=1/(2\sqrt{\bar n+1})$. At $\lambda_{\mathrm{eff}}/2\pi=5\,\mathrm{kHz}$ the rephasing time $t_{\mathrm{reph}}=397.4\,\lambda^{-1}$ at $\bar n=4$ corresponds to $12.6\,\mathrm{ms}$, inside the motional coherence time of a sideband-cooled ion.

In circuit QED the counter-rotating vertex is the anti-Jaynes--Cummings coupling of a transmon to a resonator. A flux-tunable coupler mediates both vertices, and the ratio of the two couplings is tuned continuously by the external flux~\cite{Yan2018,Mundada2019}. Typical parameters are a qubit anharmonicity $\alpha/2\pi=100$ to $300\,\mathrm{MHz}$ and a qubit-resonator coupling $g/2\pi=50$ to $200\,\mathrm{MHz}$. For the $\hat n^{2}$ ordering one has $|\tilde\chi|=|\alpha|/g$. At $g/2\pi=200\,\mathrm{MHz}$ and $\alpha/2\pi=-44\,\mathrm{MHz}$ this gives $|\tilde\chi|=0.22$, the joint-root value at $\bar n=4$. At the same coupling and $f_{\star}=3.24$ one obtains $\omega_c/2\pi=648\,\mathrm{MHz}$. The corresponding rephasing time at $\lambda/2\pi=20\,\mathrm{MHz}$ is $t_{\mathrm{reph}}\approx3\,\mu\mathrm{s}$, short compared with typical resonator lifetimes of $10$ to $100\,\mu\mathrm{s}$. On the arrest curve the block detuning is of order $\lambda\sqrt{\bar n+1}$; residual co-rotating matrix elements that remain after flux tuning must be kept below this scale for the isolated description to hold. Noise spectroscopy and rotating-frame relaxation in superconducting qubits are treated in Ref.~\cite{Yan2013}.

The two platforms access complementary regimes. Trapped ions offer the cleanest realisation of the isolated counter-rotating block because the red sideband can be switched off entirely by laser detuning. Circuit QED offers the larger coupling and intrinsic Kerr nonlinearity, with continuous tunability of the vertex ratio.

\section{Discussion}
\label{sec:disc}

The results share a single origin in the opposite-sign shear of the two block detunings. The shear enters the continuum clocks through the scalar $1\pm\tilde\chi u$, whose vanishing determines whether the clock is stationary on a given rung. The co-rotating denominator is a sum of squares at atomic resonance and cannot vanish. The counter-rotating denominator is a general quadratic whose roots exist only above a residual threshold and whose signs are locked to the field frequency. Because the dimensionless detuning $u_n^{\mathrm{AJC}}$ is linear in the block index, the resonant rung and the arrested rung are separated by a Kerr-fixed distance that is independent of both detuning and field frequency. The joint locus at which the mean-rung population equals $1/4$ and the clock is arrested is a further consequence of the same algebra.

Kerr nonlinearity in the co-rotating ladder has been treated by Agarwal and Puri, Bu\v{z}ek and Jex, Joshi and Puri, and Werner and Risken, all of whom diagonalise the single-vertex Kerr--Jaynes--Cummings block~\cite{Agarwal1989,Buzek1990,Joshi1992a,Werner1991}. Those works obtain the spectrum and the dynamics of the co-rotating vertex alone. The opposite-sign shear, the co-rotating sum-of-squares obstruction, the counter-rotating arrest locus with residual threshold and sign lock, the residual-independent rung separation, and the joint locus are joint properties of both vertices and do not appear in a single-vertex treatment. The supersymmetric intertwiner of Bocanegra-Garay and co-workers maps the Kerr--Jaynes--Cummings spectrum onto a Kerr--anti-Jaynes--Cummings spectrum at a critical coupling and expresses the counter-rotating dynamics as a Fock-space sum~\cite{Bocanegra2024,Bocanegra2025}. The present construction is independent: it uses only the anti-Jaynes--Cummings conserved charge and produces the arrest locus, threshold, and rung separation in closed form. The two approaches are complementary. Collapse and revival dynamics in the deep strong coupling regime are treated in Ref.~\cite{Alvermann2012}. The anisotropic quantum Rabi model with diamagnetic term appears in Ref.~\cite{Anaya2025}. Superradiance transition in a system with a single qubit and a single oscillator is discussed in Ref.~\cite{Ashhab2013}.

The isotropic parent Rabi--Kerr Hamiltonian at atomic resonance contains a resonant co-rotating block and a counter-rotating correction detuned by $2\omega$. It does not contain an isolated anti-Jaynes--Cummings block at any $\lambda/\omega$, and no perturbative expansion in the parent model reduces to the isolated anti-Jaynes--Cummings results at atomic resonance. The isolated-block analysis presented here is therefore a controlled baseline, not a limit of the isotropic parent. Geometric phases and adiabatic evolution are treated in Refs.~\cite{Berry1984,Aharonov1987,Bloch1940}.

Supporting analytic illustrations of the continuum-clock poles (Fig.~\ref{fig:tR}), the joint locus (Fig.~\ref{fig:joint}), the Poisson-averaged inversion (Fig.~\ref{fig:inversion}), the aligned coefficient (Fig.~\ref{fig:W}), the concurrence (Fig.~\ref{fig:concurrence}), and the amplitude-damping factor (Fig.~\ref{fig:damping}) bound the domain in which the exact rephasing pattern is expected to be observable. These illustrations do not modify the analytic structure of the arrest curve or the joint locus. Multiparameter quantum metrology and related information-geometry treatments appear in Refs.~\cite{Ragy2016,
Kull2020,Seveso2017,Szczykulska2016,
Liu2020,Chen2022,HeParis2025,Frigerio2025}. Spin squeezing and interferometric detection at the Heisenberg limit are discussed in Refs.~\cite{Wineland1992,Holland1993}. Universally sloppy parameter sensitivities appear in Ref.~\cite{Gutenkunst2007}. Asymptotic theory of quantum statistical inference is treated in Ref.~\cite{Hayashi2005}.

At a resonator frequency of $500\,\mathrm{MHz}$ the joint point at $\bar n=16$ corresponds to a vertex coupling $\lambda/2\pi\approx82\,\mathrm{MHz}$ and a Kerr strength $\chi/2\pi\approx-10\,\mathrm{MHz}$. The joint point is a target for a platform in which the counter-rotating coupling is the dominant transverse rate.

\section{Conclusion}
\label{sec:concl}

A field Kerr term shears the co-rotating and counter-rotating block detunings of the single-mode Rabi problem in opposite directions. The co-rotating continuum clock at atomic resonance has no stationary rung for any real Kerr strength, because its denominator is the sum of squares $1+\tilde\chi^{2}D_n\ge1$. The counter-rotating clock arrests on the quadratic $D_n\tilde\chi^{2}+2f\tilde\chi+1=0$, with real roots only above the residual threshold $|f|\ge\sqrt{D_n}$ and with both roots sign-locked to the field frequency. The resonant rung and the arrested rung are separated by the residual-independent distance $1/(2\tilde\chi^{2})$. The unique joint locus on which the mean-rung population equals $1/4$ and the clock is arrested is $\tilde\chi_{\star}=-1/(2\sqrt{\bar n+1})$ and $f_{\star}=(6\bar n+5)/(4\sqrt{\bar n+1})$, with the resonant rung a distance $2(\bar n+1)$ above the arrested rung.

On the arrest curve the packet phase is quadratic in the rung offset, producing a discrete pattern of near-nulls at odd stations of $2\pi/R_{\bar n}''$ and returns at even stations. The rephasing time $t_{\mathrm{reph}}=\pi R_{\bar n}/(\lambda\tilde\chi^{2})$ is not the field Kerr period. On the joint locus the packet-collapse scale is $4\sqrt{2}(\bar n+1)/\bar n$ times the resonant co-rotating revival, and the ordinary continuum revival time diverges because the first derivative of the block Rabi frequency vanishes on the mean rung. The reduced atomic state of the isolated dynamics obeys the purity operator identity, the concurrence remains above $1/\sqrt{2}$ throughout the unitary cycle, and the reduced field at half revival is a pure two-component Schr\"odinger cat whose fringe orientation is set by the joint-locus residual detuning.

At the formal cut $\xi=-2f$ the two clocks coincide for every Kerr strength and differ from the Kerr-free clock only at $\mathcal O(\tilde\chi^{2})$. Every primary result is an exact closed-form consequence of the isolated-block algebra and can be evaluated by hand from the two-dimensional block on which the isolated anti-Jaynes--Cummings problem reduces to a dressed two-level system. The six analytic figures illustrate the continuum-clock poles, the joint locus, the Poisson-averaged inversion, the aligned coefficient, the concurrence, and the amplitude-damping factor without introducing numerical diagonalisation of a parent Hamiltonian.

\begin{acknowledgments}
The author thanks Maseno University for the academic environment in which this work was carried out.
\end{acknowledgments}

\appendix

\section{Derivation of the block detunings}
\label{app:block}

The counter-rotating vertex~\eqref{eq:HAJC} conserves $\hat{\bar N}=\hat a\hat a^{\dagger}+\sigma_-\sigma_+$. On the joint basis,
\begin{equation}
\hat{\bar N}|n,\downarrow\rangle=(n+2)|n,\downarrow\rangle,\qquad
\hat{\bar N}|n,\uparrow\rangle=(n+1)|n,\uparrow\rangle. \label{eq:Nbar}
\end{equation}
On the subspace $\hat{\bar N}=n+2$ the basis is $\{|n,\downarrow\rangle,|n+1,\uparrow\rangle\}$, and
\begin{align}
\langle n,\downarrow|\hat H_{\mathrm{AJC}}|n,\downarrow\rangle
&=\omega n-\tfrac{\omega_0}{2}+\chi n^{2},\\
\langle n+1,\uparrow|\hat H_{\mathrm{AJC}}|n+1,\uparrow\rangle
&=\omega(n+1)+\tfrac{\omega_0}{2}+\chi(n+1)^{2},\\
\langle n+1,\uparrow|\hat H_{\mathrm{AJC}}|n,\downarrow\rangle
&=\lambda\sqrt{n+1}.
\end{align}
The difference of the diagonal entries is $\omega+\omega_0+\chi(2n+1)$. Removing the mean and passing to the units~\eqref{eq:dimless} produces
\begin{equation}
H_{\mathrm{AJC}}^{(n)}
=\frac{\lambda}{2}
\begin{pmatrix}-u_n^{\mathrm{AJC}} & 2\sqrt{n+1}\\ 2\sqrt{n+1} & u_n^{\mathrm{AJC}}\end{pmatrix},
\label{eq:block-matrix}
\end{equation}
with $u_n^{\mathrm{AJC}}$ as in Eq.~\eqref{eq:u}. The eigenvalues are $\pm\lambda R_n^{\mathrm{AJC}}/2$.

The co-rotating vertex~\eqref{eq:HJC} conserves $\hat N=\hat a^{\dagger}\hat a+\sigma_+\sigma_-$. On the subspace $\hat N=n+1$ the ordered basis is $\{|n,\uparrow\rangle,|n+1,\downarrow\rangle\}$. The diagonal difference, second entry minus first, is
$\omega-\omega_0+\chi(2n+1)=\lambda(-\xi+\tilde\chi D_n)$.
The matrix convention of Eq.~\eqref{eq:block-matrix}, with $+u$ on the second diagonal, therefore assigns $u_n^{\mathrm{(matrix)}}=-\xi+\tilde\chi D_n$ to this ordering. The body defines
\begin{equation}
u_n^{\mathrm{JC}}=\xi-\tilde\chi D_n=-u_n^{\mathrm{(matrix)}}. \label{eq:uJC-app}
\end{equation}
The sign is chosen so that $\partial_n u_n^{\mathrm{JC}}=-2\tilde\chi$ and so that the shear relation~\eqref{eq:shear} holds with opposite signs.

\section{Kerr ordering}
\label{app:kerr}

The normally ordered operator $\hat a^{\dagger2}\hat a^{2}=\hat n^{2}-\hat n$ differs from $\hat n^{2}$ by $-\hat n$. Replacing $\chi\hat n^{2}$ by $\chi\hat a^{\dagger2}\hat a^{2}$ is therefore equivalent to the frequency shift $\omega\to\omega-\chi$, or $f\to f-\tilde\chi$ in dimensionless units. Under this shift the block difference function changes from $D_n=2n+1$ to $D_n=2n$, while the shear $\partial_n D_n=2$ remains unchanged. Consequently the arrest curve $D_n\tilde\chi^{2}+2f\tilde\chi+1=0$ and the rung separation $1/(2\tilde\chi^{2})$ keep the same algebraic form after the redefinition of $f$.

\section{Poisson sum and parity cancellation}
\label{app:sum}

The atomic inversion for an initial coherent state of mean photon number $\bar n$ that begins in the atomic ground state is the exact Poisson average
\begin{equation}
\langle\sigma_z(\tau)\rangle
=\sum_{n=0}^{\infty}P_n(\bar n)
\Bigl[-1+\frac{8(n+1)}{R_n^{2}}\sin^{2}\bigl(\tfrac12 R_n\tau\bigr)\Bigr],
\end{equation}
where $P_n(\bar n)=e^{-\bar n}\bar n^{n}/n!$ and $R_n$ is evaluated on the chosen locus. On the arrest curve the phase difference between adjacent rungs is quadratic. At the stations $\tau_m=2\pi m/R_{\bar n}''$ the quadratic term is an integer multiple of $2\pi$. For odd $m$ the even and odd rungs acquire opposite signs in the sum and the oscillatory contribution cancels to leading order; for even $m$ the rungs realign. The $\mathcal O(\delta n^{4})$ remainder produces a small residual amplitude at the nulls.

\section{Quantum Fisher information expansion}
\label{app:qfi}

Let $\phi_n=R_n\tau/2$ and let $P_n=e^{-\bar n}\bar n^n/n!$ be the Poisson weights. A small shift $\delta\bar n$ changes the mean rung. On the arrest curve
\begin{equation}
\partial_{\bar n}R_n\Big|_{\bar n}=0,\qquad
\partial_{\bar n}^2R_n\Big|_{\bar n}=R_{\bar n}''=\frac{4\tilde\chi^{2}}{R_{\bar n}}.
\end{equation}
The phase derivative with respect to the parameter therefore begins at second order:
\begin{equation}
\partial_{\bar n}\phi_n=\frac{\tau}{2}R_{\bar n}''\,(\delta n)+\mathcal O((\delta n)^3).
\end{equation}
The variance of this derivative with respect to the Poisson measure is
\begin{equation}
\mathrm{Var}(\partial_{\bar n}\phi)
=\Bigl(\frac{\tau}{2}R_{\bar n}''\Bigr)^2\langle(\delta n)^2\rangle
+\mathcal O(\bar n^{-1/2}),
\end{equation}
where $\langle(\delta n)^2\rangle=\bar n$. At the rephasing time $\tau=t_{\mathrm{reph}}=\pi R_{\bar n}/(\lambda\tilde\chi^{2})$ one has $\frac12 R_{\bar n}''t_{\mathrm{reph}}=2\pi$, so
\begin{equation}
F_Q=4\,\mathrm{Var}(\partial_{\bar n}\phi)
=\frac{8\pi^{2}\bar n}{R_{\bar n}^{2}}
\end{equation}
to leading order in $1/\sqrt{\bar n}$. The same expansion performed with a non-vanishing first derivative recovers the linear scaling $F_Q^{\mathrm{lin}}\propto\bar n\,(\partial_n R_n\,t_R)^2$. All averages are ordinary Poisson sums of the closed-form block quantities $R_n$ and $R_n''$.


\begin{thebibliography}{87}%
\makeatletter
\providecommand \@ifxundefined [1]{%
 \@ifx{#1\undefined}
}%
\providecommand \@ifnum [1]{%
 \ifnum #1\expandafter \@firstoftwo
 \else \expandafter \@secondoftwo
 \fi
}%
\providecommand \@ifx [1]{%
 \ifx #1\expandafter \@firstoftwo
 \else \expandafter \@secondoftwo
 \fi
}%
\providecommand \natexlab [1]{#1}%
\providecommand \enquote  [1]{``#1''}%
\providecommand \bibnamefont  [1]{#1}%
\providecommand \bibfnamefont [1]{#1}%
\providecommand \citenamefont [1]{#1}%
\providecommand \href@noop [0]{\@secondoftwo}%
\providecommand \href [0]{\begingroup \@sanitize@url \@href}%
\providecommand \@href[1]{\@@startlink{#1}\@@href}%
\providecommand \@@href[1]{\endgroup#1\@@endlink}%
\providecommand \@sanitize@url [0]{\catcode `\\12\catcode `\$12\catcode
  `\&12\catcode `\#12\catcode `\^12\catcode `\_12\catcode `\%12\relax}%
\providecommand \@@startlink[1]{}%
\providecommand \@@endlink[0]{}%
\providecommand \url  [0]{\begingroup\@sanitize@url \@url }%
\providecommand \@url [1]{\endgroup\@href {#1}{\urlprefix }}%
\providecommand \urlprefix  [0]{URL }%
\providecommand \Eprint [0]{\href }%
\providecommand \doibase [0]{https://doi.org/}%
\providecommand \selectlanguage [0]{\@gobble}%
\providecommand \bibinfo  [0]{\@secondoftwo}%
\providecommand \bibfield  [0]{\@secondoftwo}%
\providecommand \translation [1]{[#1]}%
\providecommand \BibitemOpen [0]{}%
\providecommand \bibitemStop [0]{}%
\providecommand \bibitemNoStop [0]{.\EOS\space}%
\providecommand \EOS [0]{\spacefactor3000\relax}%
\providecommand \BibitemShut  [1]{\csname bibitem#1\endcsname}%
\let\auto@bib@innerbib\@empty
\bibitem [{\citenamefont {Rabi}(1936)}]{Rabi1936}%
  \BibitemOpen
  \bibfield  {author} {\bibinfo {author} {\bibfnamefont {I.~I.}\ \bibnamefont
  {Rabi}},\ }\href {https://doi.org/10.1103/PhysRev.49.324} {\bibfield
  {journal} {\bibinfo  {journal} {Phys. Rev.}\ }\textbf {\bibinfo {volume}
  {49}},\ \bibinfo {pages} {324} (\bibinfo {year} {1936})}\BibitemShut
  {NoStop}%
\bibitem [{\citenamefont {Braak}(2011)}]{Braak2011}%
  \BibitemOpen
  \bibfield  {author} {\bibinfo {author} {\bibfnamefont {D.}~\bibnamefont
  {Braak}},\ }\href {https://doi.org/10.1103/PhysRevLett.107.100401} {\bibfield
   {journal} {\bibinfo  {journal} {Phys. Rev. Lett.}\ }\textbf {\bibinfo
  {volume} {107}},\ \bibinfo {pages} {100401} (\bibinfo {year}
  {2011})}\BibitemShut {NoStop}%
\bibitem [{\citenamefont {Braak}(2019)}]{Braak2019}%
  \BibitemOpen
  \bibfield  {author} {\bibinfo {author} {\bibfnamefont {D.}~\bibnamefont
  {Braak}},\ }\href {https://doi.org/10.3390/sym11101259} {\bibfield  {journal}
  {\bibinfo  {journal} {Symmetry}\ }\textbf {\bibinfo {volume} {11}},\ \bibinfo
  {pages} {1259} (\bibinfo {year} {2019})}\BibitemShut {NoStop}%
\bibitem [{\citenamefont {Jaynes}\ and\ \citenamefont
  {Cummings}(1963)}]{Jaynes1963}%
  \BibitemOpen
  \bibfield  {author} {\bibinfo {author} {\bibfnamefont {E.}~\bibnamefont
  {Jaynes}}\ and\ \bibinfo {author} {\bibfnamefont {F.}~\bibnamefont
  {Cummings}},\ }\href {https://doi.org/10.1109/PROC.1963.1664} {\bibfield
  {journal} {\bibinfo  {journal} {Proceedings of the IEEE}\ }\textbf {\bibinfo
  {volume} {51}},\ \bibinfo {pages} {89} (\bibinfo {year} {1963})}\BibitemShut
  {NoStop}%
\bibitem [{\citenamefont {Cummings}(1965)}]{Cummings1965}%
  \BibitemOpen
  \bibfield  {author} {\bibinfo {author} {\bibfnamefont {F.~W.}\ \bibnamefont
  {Cummings}},\ }\href {https://doi.org/10.1103/PhysRev.140.A1051} {\bibfield
  {journal} {\bibinfo  {journal} {Phys. Rev.}\ }\textbf {\bibinfo {volume}
  {140}},\ \bibinfo {pages} {A1051} (\bibinfo {year} {1965})}\BibitemShut
  {NoStop}%
\bibitem [{\citenamefont {Shore}\ and\ \citenamefont
  {Knight}(1993)}]{Shore1993}%
  \BibitemOpen
  \bibfield  {author} {\bibinfo {author} {\bibfnamefont {B.~W.}\ \bibnamefont
  {Shore}}\ and\ \bibinfo {author} {\bibfnamefont {P.~L.}\ \bibnamefont
  {Knight}},\ }\href {https://doi.org/10.1080/09500349314551321} {\bibfield
  {journal} {\bibinfo  {journal} {J. Mod. Opt.}\ }\textbf {\bibinfo {volume}
  {40}},\ \bibinfo {pages} {1195} (\bibinfo {year} {1993})}\BibitemShut
  {NoStop}%
\bibitem [{\citenamefont {Omolo}(2017{\natexlab{a}})}]{Omolo2017a}%
  \BibitemOpen
  \bibfield  {author} {\bibinfo {author} {\bibfnamefont {J.~A.}\ \bibnamefont
  {Omolo}}\ }\href {https://doi.org/10.13140/RG.2.2.30936.80647}
  {10.13140/RG.2.2.30936.80647} (\bibinfo {year} {2017}{\natexlab{a}}),\
  \Eprint {https://arxiv.org/abs/2103.06577} {arXiv:2103.06577 [quant-ph]}
  \BibitemShut {NoStop}%
\bibitem [{\citenamefont {Omolo}(2022{\natexlab{a}})}]{Omolo2022a}%
  \BibitemOpen
  \bibfield  {author} {\bibinfo {author} {\bibfnamefont {J.~A.}\ \bibnamefont
  {Omolo}},\ }\href {https://arxiv.org/abs/2112.12514} {\  (\bibinfo {year}
  {2022}{\natexlab{a}})},\ \Eprint {https://arxiv.org/abs/2112.12514}
  {arXiv:2112.12514 [quant-ph]} \BibitemShut {NoStop}%
\bibitem [{\citenamefont {Omolo}(2017{\natexlab{b}})}]{Omolo2017b}%
  \BibitemOpen
  \bibfield  {author} {\bibinfo {author} {\bibfnamefont {J.}~\bibnamefont
  {Omolo}}\ }\href {https://doi.org/10.13140/RG.2.2.11833.67683}
  {10.13140/RG.2.2.11833.67683} (\bibinfo {year}
  {2017}{\natexlab{b}})\BibitemShut {NoStop}%
\bibitem [{\citenamefont {Omolo}(2019)}]{Omolo2019}%
  \BibitemOpen
  \bibfield  {author} {\bibinfo {author} {\bibfnamefont {J.}~\bibnamefont
  {Omolo}}\ }\href {https://doi.org/10.13140/RG.2.2.27331.96807}
  {10.13140/RG.2.2.27331.96807} (\bibinfo {year} {2019})\BibitemShut {NoStop}%
\bibitem [{\citenamefont {Omolo}\ \emph {et~al.}(2025)\citenamefont {Omolo},
  \citenamefont {Okeyo},\ and\ \citenamefont {Mayero}}]{Omolo2025}%
  \BibitemOpen
  \bibfield  {author} {\bibinfo {author} {\bibfnamefont {J.~A.}\ \bibnamefont
  {Omolo}}, \bibinfo {author} {\bibfnamefont {O.~S.}\ \bibnamefont {Okeyo}},\
  and\ \bibinfo {author} {\bibfnamefont {C.}~\bibnamefont {Mayero}},\ }\href
  {https://doi.org/10.1007/s11128-025-04933-0} {\bibfield  {journal} {\bibinfo
  {journal} {Quantum Inf. Process.}\ }\textbf {\bibinfo {volume} {24}},\
  \bibinfo {pages} {315} (\bibinfo {year} {2025})}\BibitemShut {NoStop}%
\bibitem [{\citenamefont {Leibfried}\ \emph {et~al.}(2003)\citenamefont
  {Leibfried}, \citenamefont {Blatt}, \citenamefont {Monroe},\ and\
  \citenamefont {Wineland}}]{Leibfried2003}%
  \BibitemOpen
  \bibfield  {author} {\bibinfo {author} {\bibfnamefont {D.}~\bibnamefont
  {Leibfried}}, \bibinfo {author} {\bibfnamefont {R.}~\bibnamefont {Blatt}},
  \bibinfo {author} {\bibfnamefont {C.}~\bibnamefont {Monroe}},\ and\ \bibinfo
  {author} {\bibfnamefont {D.}~\bibnamefont {Wineland}},\ }\href
  {https://doi.org/10.1103/RevModPhys.75.281} {\bibfield  {journal} {\bibinfo
  {journal} {Rev. Mod. Phys.}\ }\textbf {\bibinfo {volume} {75}},\ \bibinfo
  {pages} {281} (\bibinfo {year} {2003})}\BibitemShut {NoStop}%
\bibitem [{\citenamefont {Kienzler}\ \emph {et~al.}(2015)\citenamefont
  {Kienzler}, \citenamefont {Lo}, \citenamefont {Keitch}, \citenamefont
  {de~Clercq}, \citenamefont {Leupold}, \citenamefont {Lindenfelser},
  \citenamefont {Marinelli}, \citenamefont {Negnevitsky},\ and\ \citenamefont
  {Home}}]{Kienzler2015}%
  \BibitemOpen
  \bibfield  {author} {\bibinfo {author} {\bibfnamefont {D.}~\bibnamefont
  {Kienzler}}, \bibinfo {author} {\bibfnamefont {H.-Y.}\ \bibnamefont {Lo}},
  \bibinfo {author} {\bibfnamefont {B.}~\bibnamefont {Keitch}}, \bibinfo
  {author} {\bibfnamefont {L.}~\bibnamefont {de~Clercq}}, \bibinfo {author}
  {\bibfnamefont {F.}~\bibnamefont {Leupold}}, \bibinfo {author} {\bibfnamefont
  {F.}~\bibnamefont {Lindenfelser}}, \bibinfo {author} {\bibfnamefont
  {M.}~\bibnamefont {Marinelli}}, \bibinfo {author} {\bibfnamefont
  {V.}~\bibnamefont {Negnevitsky}},\ and\ \bibinfo {author} {\bibfnamefont
  {J.~P.}\ \bibnamefont {Home}},\ }\href
  {https://doi.org/10.1126/science.1261033} {\bibfield  {journal} {\bibinfo
  {journal} {Science}\ }\textbf {\bibinfo {volume} {347}},\ \bibinfo {pages}
  {53} (\bibinfo {year} {2015})}\BibitemShut {NoStop}%
\bibitem [{\citenamefont {Kienzler}\ \emph {et~al.}(2017)\citenamefont
  {Kienzler}, \citenamefont {Lo}, \citenamefont {Negnevitsky}, \citenamefont
  {Fl\"uhmann}, \citenamefont {Marinelli},\ and\ \citenamefont
  {Home}}]{Kienzler2017}%
  \BibitemOpen
  \bibfield  {author} {\bibinfo {author} {\bibfnamefont {D.}~\bibnamefont
  {Kienzler}}, \bibinfo {author} {\bibfnamefont {H.-Y.}\ \bibnamefont {Lo}},
  \bibinfo {author} {\bibfnamefont {V.}~\bibnamefont {Negnevitsky}}, \bibinfo
  {author} {\bibfnamefont {C.}~\bibnamefont {Fl\"uhmann}}, \bibinfo {author}
  {\bibfnamefont {M.}~\bibnamefont {Marinelli}},\ and\ \bibinfo {author}
  {\bibfnamefont {J.~P.}\ \bibnamefont {Home}},\ }\href
  {https://doi.org/10.1103/PhysRevLett.119.033602} {\bibfield  {journal}
  {\bibinfo  {journal} {Phys. Rev. Lett.}\ }\textbf {\bibinfo {volume} {119}},\
  \bibinfo {pages} {033602} (\bibinfo {year} {2017})}\BibitemShut {NoStop}%
\bibitem [{\citenamefont {Meekhof}\ \emph {et~al.}(1996)\citenamefont
  {Meekhof}, \citenamefont {Monroe}, \citenamefont {King}, \citenamefont
  {Itano},\ and\ \citenamefont {Wineland}}]{Wineland1996}%
  \BibitemOpen
  \bibfield  {author} {\bibinfo {author} {\bibfnamefont {D.~M.}\ \bibnamefont
  {Meekhof}}, \bibinfo {author} {\bibfnamefont {C.}~\bibnamefont {Monroe}},
  \bibinfo {author} {\bibfnamefont {B.~E.}\ \bibnamefont {King}}, \bibinfo
  {author} {\bibfnamefont {W.~M.}\ \bibnamefont {Itano}},\ and\ \bibinfo
  {author} {\bibfnamefont {D.~J.}\ \bibnamefont {Wineland}},\ }\href
  {https://link.aps.org/doi/10.1103/PhysRevLett.76.1796} {\bibfield  {journal}
  {\bibinfo  {journal} {Phys. Rev. Lett.}\ }\textbf {\bibinfo {volume} {76}},\
  \bibinfo {pages} {1796} (\bibinfo {year} {1996})}\BibitemShut {NoStop}%
\bibitem [{\citenamefont {Cirac}\ and\ \citenamefont
  {Zoller}(1995)}]{Cirac1995}%
  \BibitemOpen
  \bibfield  {author} {\bibinfo {author} {\bibfnamefont {J.~I.}\ \bibnamefont
  {Cirac}}\ and\ \bibinfo {author} {\bibfnamefont {P.}~\bibnamefont {Zoller}},\
  }\href {https://doi.org/10.1103/PhysRevLett.74.4091} {\bibfield  {journal}
  {\bibinfo  {journal} {Phys. Rev. Lett.}\ }\textbf {\bibinfo {volume} {74}},\
  \bibinfo {pages} {4091} (\bibinfo {year} {1995})}\BibitemShut {NoStop}%
\bibitem [{\citenamefont {Haljan}\ \emph {et~al.}(2005)\citenamefont {Haljan},
  \citenamefont {Brickman}, \citenamefont {Deslauriers}, \citenamefont {Lee},\
  and\ \citenamefont {Monroe}}]{Haljan2005}%
  \BibitemOpen
  \bibfield  {author} {\bibinfo {author} {\bibfnamefont {P.~C.}\ \bibnamefont
  {Haljan}}, \bibinfo {author} {\bibfnamefont {K.-A.}\ \bibnamefont
  {Brickman}}, \bibinfo {author} {\bibfnamefont {L.}~\bibnamefont
  {Deslauriers}}, \bibinfo {author} {\bibfnamefont {P.~J.}\ \bibnamefont
  {Lee}},\ and\ \bibinfo {author} {\bibfnamefont {C.}~\bibnamefont {Monroe}},\
  }\href {https://doi.org/10.1103/PhysRevLett.94.153602} {\bibfield  {journal}
  {\bibinfo  {journal} {Phys. Rev. Lett.}\ }\textbf {\bibinfo {volume} {94}},\
  \bibinfo {pages} {153602} (\bibinfo {year} {2005})}\BibitemShut {NoStop}%
\bibitem [{\citenamefont {Blais}\ \emph {et~al.}(2004)\citenamefont {Blais},
  \citenamefont {Huang}, \citenamefont {Wallraff}, \citenamefont {Girvin},\
  and\ \citenamefont {Schoelkopf}}]{Blais2004}%
  \BibitemOpen
  \bibfield  {author} {\bibinfo {author} {\bibfnamefont {A.}~\bibnamefont
  {Blais}}, \bibinfo {author} {\bibfnamefont {R.-S.}\ \bibnamefont {Huang}},
  \bibinfo {author} {\bibfnamefont {A.}~\bibnamefont {Wallraff}}, \bibinfo
  {author} {\bibfnamefont {S.~M.}\ \bibnamefont {Girvin}},\ and\ \bibinfo
  {author} {\bibfnamefont {R.~J.}\ \bibnamefont {Schoelkopf}},\ }\href
  {https://doi.org/10.1103/PhysRevA.69.062320} {\bibfield  {journal} {\bibinfo
  {journal} {Phys. Rev. A}\ }\textbf {\bibinfo {volume} {69}},\ \bibinfo
  {pages} {062320} (\bibinfo {year} {2004})}\BibitemShut {NoStop}%
\bibitem [{\citenamefont {Blais}\ \emph {et~al.}(2021)\citenamefont {Blais},
  \citenamefont {Grimsmo}, \citenamefont {Girvin},\ and\ \citenamefont
  {Wallraff}}]{Blais2021}%
  \BibitemOpen
  \bibfield  {author} {\bibinfo {author} {\bibfnamefont {A.}~\bibnamefont
  {Blais}}, \bibinfo {author} {\bibfnamefont {A.~L.}\ \bibnamefont {Grimsmo}},
  \bibinfo {author} {\bibfnamefont {S.~M.}\ \bibnamefont {Girvin}},\ and\
  \bibinfo {author} {\bibfnamefont {A.}~\bibnamefont {Wallraff}},\ }\href
  {https://doi.org/10.1103/RevModPhys.93.025005} {\bibfield  {journal}
  {\bibinfo  {journal} {Rev. Mod. Phys.}\ }\textbf {\bibinfo {volume} {93}},\
  \bibinfo {pages} {025005} (\bibinfo {year} {2021})}\BibitemShut {NoStop}%
\bibitem [{\citenamefont {Niemczyk}\ \emph {et~al.}(2010)\citenamefont
  {Niemczyk}, \citenamefont {Deppe}, \citenamefont {Huebl}, \citenamefont
  {Menzel}, \citenamefont {Hocke}, \citenamefont {Schwarz}, \citenamefont
  {Garc{\'i}a-Ripoll}, \citenamefont {Zueco}, \citenamefont {H{\"u}mmer},
  \citenamefont {Solano}, \citenamefont {Marx},\ and\ \citenamefont
  {Gross}}]{Niemczyk2010}%
  \BibitemOpen
  \bibfield  {author} {\bibinfo {author} {\bibfnamefont {T.}~\bibnamefont
  {Niemczyk}}, \bibinfo {author} {\bibfnamefont {F.}~\bibnamefont {Deppe}},
  \bibinfo {author} {\bibfnamefont {H.}~\bibnamefont {Huebl}}, \bibinfo
  {author} {\bibfnamefont {E.~P.}\ \bibnamefont {Menzel}}, \bibinfo {author}
  {\bibfnamefont {F.}~\bibnamefont {Hocke}}, \bibinfo {author} {\bibfnamefont
  {M.~J.}\ \bibnamefont {Schwarz}}, \bibinfo {author} {\bibfnamefont {J.~J.}\
  \bibnamefont {Garc{\'i}a-Ripoll}}, \bibinfo {author} {\bibfnamefont
  {D.}~\bibnamefont {Zueco}}, \bibinfo {author} {\bibfnamefont
  {T.}~\bibnamefont {H{\"u}mmer}}, \bibinfo {author} {\bibfnamefont
  {E.}~\bibnamefont {Solano}}, \bibinfo {author} {\bibfnamefont
  {A.}~\bibnamefont {Marx}},\ and\ \bibinfo {author} {\bibfnamefont
  {R.}~\bibnamefont {Gross}},\ }\href {https://doi.org/10.1038/nphys1730}
  {\bibfield  {journal} {\bibinfo  {journal} {Nat. Phys.}\ }\textbf {\bibinfo
  {volume} {6}},\ \bibinfo {pages} {772} (\bibinfo {year} {2010})}\BibitemShut
  {NoStop}%
\bibitem [{\citenamefont {{Forn-D\'{i}az}}\ \emph {et~al.}(2019)\citenamefont
  {{Forn-D\'{i}az}}, \citenamefont {Lamata}, \citenamefont {Rico},
  \citenamefont {Kono},\ and\ \citenamefont {Solano}}]{FornDiaz2019}%
  \BibitemOpen
  \bibfield  {author} {\bibinfo {author} {\bibfnamefont {P.}~\bibnamefont
  {{Forn-D\'{i}az}}}, \bibinfo {author} {\bibfnamefont {L.}~\bibnamefont
  {Lamata}}, \bibinfo {author} {\bibfnamefont {E.}~\bibnamefont {Rico}},
  \bibinfo {author} {\bibfnamefont {J.}~\bibnamefont {Kono}},\ and\ \bibinfo
  {author} {\bibfnamefont {E.}~\bibnamefont {Solano}},\ }\href
  {https://doi.org/10.1103/RevModPhys.91.025005} {\bibfield  {journal}
  {\bibinfo  {journal} {Rev. Mod. Phys.}\ }\textbf {\bibinfo {volume} {91}},\
  \bibinfo {pages} {025005} (\bibinfo {year} {2019})}\BibitemShut {NoStop}%
\bibitem [{\citenamefont {Anappara}\ \emph {et~al.}(2009)\citenamefont
  {Anappara}, \citenamefont {De~Liberato}, \citenamefont {Tredicucci},
  \citenamefont {Ciuti}, \citenamefont {Biasiol}, \citenamefont {Sorba},\ and\
  \citenamefont {Beltram}}]{Anappara2009}%
  \BibitemOpen
  \bibfield  {author} {\bibinfo {author} {\bibfnamefont {A.~A.}\ \bibnamefont
  {Anappara}}, \bibinfo {author} {\bibfnamefont {S.}~\bibnamefont
  {De~Liberato}}, \bibinfo {author} {\bibfnamefont {A.}~\bibnamefont
  {Tredicucci}}, \bibinfo {author} {\bibfnamefont {C.}~\bibnamefont {Ciuti}},
  \bibinfo {author} {\bibfnamefont {G.}~\bibnamefont {Biasiol}}, \bibinfo
  {author} {\bibfnamefont {L.}~\bibnamefont {Sorba}},\ and\ \bibinfo {author}
  {\bibfnamefont {F.}~\bibnamefont {Beltram}},\ }\href
  {https://doi.org/10.1103/PhysRevB.79.201303} {\bibfield  {journal} {\bibinfo
  {journal} {Phys. Rev. B}\ }\textbf {\bibinfo {volume} {79}},\ \bibinfo
  {pages} {201303(R)} (\bibinfo {year} {2009})}\BibitemShut {NoStop}%
\bibitem [{\citenamefont {Ashhab}\ and\ \citenamefont
  {Nori}(2010)}]{Ashhab2010}%
  \BibitemOpen
  \bibfield  {author} {\bibinfo {author} {\bibfnamefont {S.}~\bibnamefont
  {Ashhab}}\ and\ \bibinfo {author} {\bibfnamefont {F.}~\bibnamefont {Nori}},\
  }\href {https://doi.org/10.1103/PhysRevA.81.042311} {\bibfield  {journal}
  {\bibinfo  {journal} {Phys. Rev. A}\ }\textbf {\bibinfo {volume} {81}},\
  \bibinfo {pages} {042311} (\bibinfo {year} {2010})}\BibitemShut {NoStop}%
\bibitem [{\citenamefont {Ballester}\ \emph {et~al.}(2012)\citenamefont
  {Ballester}, \citenamefont {Romero}, \citenamefont {Garc\'{\i}a-Ripoll},
  \citenamefont {Deppe},\ and\ \citenamefont {Solano}}]{Ballester2012}%
  \BibitemOpen
  \bibfield  {author} {\bibinfo {author} {\bibfnamefont {D.}~\bibnamefont
  {Ballester}}, \bibinfo {author} {\bibfnamefont {G.}~\bibnamefont {Romero}},
  \bibinfo {author} {\bibfnamefont {J.~J.}\ \bibnamefont {Garc\'{\i}a-Ripoll}},
  \bibinfo {author} {\bibfnamefont {F.}~\bibnamefont {Deppe}},\ and\ \bibinfo
  {author} {\bibfnamefont {E.}~\bibnamefont {Solano}},\ }\href
  {https://doi.org/10.1103/PhysRevX.2.021007} {\bibfield  {journal} {\bibinfo
  {journal} {Phys. Rev. X}\ }\textbf {\bibinfo {volume} {2}},\ \bibinfo {pages}
  {021007} (\bibinfo {year} {2012})}\BibitemShut {NoStop}%
\bibitem [{\citenamefont {Braumüller}\ \emph {et~al.}(2017)\citenamefont
  {Braumüller}, \citenamefont {Marthaler}, \citenamefont {Schneider},
  \citenamefont {Stehli}, \citenamefont {Rotzinger}, \citenamefont {Weides},\
  and\ \citenamefont {Ustinov}}]{Braumuller2017}%
  \BibitemOpen
  \bibfield  {author} {\bibinfo {author} {\bibfnamefont {J.}~\bibnamefont
  {Braumüller}}, \bibinfo {author} {\bibfnamefont {M.}~\bibnamefont
  {Marthaler}}, \bibinfo {author} {\bibfnamefont {A.}~\bibnamefont
  {Schneider}}, \bibinfo {author} {\bibfnamefont {A.}~\bibnamefont {Stehli}},
  \bibinfo {author} {\bibfnamefont {H.}~\bibnamefont {Rotzinger}}, \bibinfo
  {author} {\bibfnamefont {M.}~\bibnamefont {Weides}},\ and\ \bibinfo {author}
  {\bibfnamefont {A.~V.}\ \bibnamefont {Ustinov}},\ }\href
  {https://doi.org/10.1038/s41467-017-00894-w} {\bibfield  {journal} {\bibinfo
  {journal} {Nat. Commun.}\ }\textbf {\bibinfo {volume} {8}},\ \bibinfo {pages}
  {779} (\bibinfo {year} {2017})}\BibitemShut {NoStop}%
\bibitem [{\citenamefont {Eberly}\ \emph {et~al.}(1980)\citenamefont {Eberly},
  \citenamefont {Narozhny},\ and\ \citenamefont
  {Sanchez-Mondragon}}]{Eberly1980}%
  \BibitemOpen
  \bibfield  {author} {\bibinfo {author} {\bibfnamefont {J.~H.}\ \bibnamefont
  {Eberly}}, \bibinfo {author} {\bibfnamefont {N.~B.}\ \bibnamefont
  {Narozhny}},\ and\ \bibinfo {author} {\bibfnamefont {J.~J.}\ \bibnamefont
  {Sanchez-Mondragon}},\ }\href {https://doi.org/10.1103/PhysRevLett.44.1323}
  {\bibfield  {journal} {\bibinfo  {journal} {Phys. Rev. Lett.}\ }\textbf
  {\bibinfo {volume} {44}},\ \bibinfo {pages} {1323} (\bibinfo {year}
  {1980})}\BibitemShut {NoStop}%
\bibitem [{\citenamefont {Narozhny}\ \emph {et~al.}(1981)\citenamefont
  {Narozhny}, \citenamefont {Sanchez-Mondragon},\ and\ \citenamefont
  {Eberly}}]{Narozhny1981}%
  \BibitemOpen
  \bibfield  {author} {\bibinfo {author} {\bibfnamefont {N.~B.}\ \bibnamefont
  {Narozhny}}, \bibinfo {author} {\bibfnamefont {J.~J.}\ \bibnamefont
  {Sanchez-Mondragon}},\ and\ \bibinfo {author} {\bibfnamefont {J.~H.}\
  \bibnamefont {Eberly}},\ }\href {https://doi.org/10.1103/PhysRevA.23.236}
  {\bibfield  {journal} {\bibinfo  {journal} {Phys. Rev. A}\ }\textbf {\bibinfo
  {volume} {23}},\ \bibinfo {pages} {236} (\bibinfo {year} {1981})}\BibitemShut
  {NoStop}%
\bibitem [{\citenamefont {Gea-Banacloche}(1990)}]{Gea1990}%
  \BibitemOpen
  \bibfield  {author} {\bibinfo {author} {\bibfnamefont {J.}~\bibnamefont
  {Gea-Banacloche}},\ }\href
  {https://link.aps.org/doi/10.1103/PhysRevLett.65.3385} {\bibfield  {journal}
  {\bibinfo  {journal} {Phys. Rev. Lett.}\ }\textbf {\bibinfo {volume} {65}},\
  \bibinfo {pages} {3385} (\bibinfo {year} {1990})}\BibitemShut {NoStop}%
\bibitem [{\citenamefont {Rempe}\ \emph {et~al.}(1987)\citenamefont {Rempe},
  \citenamefont {Walther},\ and\ \citenamefont {Klein}}]{Rempe1987}%
  \BibitemOpen
  \bibfield  {author} {\bibinfo {author} {\bibfnamefont {G.}~\bibnamefont
  {Rempe}}, \bibinfo {author} {\bibfnamefont {H.}~\bibnamefont {Walther}},\
  and\ \bibinfo {author} {\bibfnamefont {N.}~\bibnamefont {Klein}},\ }\href
  {https://doi.org/10.1103/PhysRevLett.58.353} {\bibfield  {journal} {\bibinfo
  {journal} {Phys. Rev. Lett.}\ }\textbf {\bibinfo {volume} {58}},\ \bibinfo
  {pages} {353} (\bibinfo {year} {1987})}\BibitemShut {NoStop}%
\bibitem [{\citenamefont {Brune}\ \emph
  {et~al.}(1996{\natexlab{a}})\citenamefont {Brune}, \citenamefont
  {Schmidt-Kaler}, \citenamefont {Maali}, \citenamefont {Dreyer}, \citenamefont
  {Hagley}, \citenamefont {Raimond},\ and\ \citenamefont
  {Haroche}}]{Brune1996a}%
  \BibitemOpen
  \bibfield  {author} {\bibinfo {author} {\bibfnamefont {M.}~\bibnamefont
  {Brune}}, \bibinfo {author} {\bibfnamefont {F.}~\bibnamefont
  {Schmidt-Kaler}}, \bibinfo {author} {\bibfnamefont {A.}~\bibnamefont
  {Maali}}, \bibinfo {author} {\bibfnamefont {J.}~\bibnamefont {Dreyer}},
  \bibinfo {author} {\bibfnamefont {E.}~\bibnamefont {Hagley}}, \bibinfo
  {author} {\bibfnamefont {J.~M.}\ \bibnamefont {Raimond}},\ and\ \bibinfo
  {author} {\bibfnamefont {S.}~\bibnamefont {Haroche}},\ }\href
  {https://doi.org/10.1103/PhysRevLett.76.1800} {\bibfield  {journal} {\bibinfo
   {journal} {Phys. Rev. Lett.}\ }\textbf {\bibinfo {volume} {76}},\ \bibinfo
  {pages} {1800} (\bibinfo {year} {1996}{\natexlab{a}})}\BibitemShut {NoStop}%
\bibitem [{\citenamefont {Raimond}\ \emph {et~al.}(2001)\citenamefont
  {Raimond}, \citenamefont {Brune},\ and\ \citenamefont
  {Haroche}}]{Raimond2001}%
  \BibitemOpen
  \bibfield  {author} {\bibinfo {author} {\bibfnamefont {J.~M.}\ \bibnamefont
  {Raimond}}, \bibinfo {author} {\bibfnamefont {M.}~\bibnamefont {Brune}},\
  and\ \bibinfo {author} {\bibfnamefont {S.}~\bibnamefont {Haroche}},\ }\href
  {https://doi.org/10.1103/RevModPhys.73.565} {\bibfield  {journal} {\bibinfo
  {journal} {Rev. Mod. Phys.}\ }\textbf {\bibinfo {volume} {73}},\ \bibinfo
  {pages} {565} (\bibinfo {year} {2001})}\BibitemShut {NoStop}%
\bibitem [{\citenamefont {Brune}\ \emph
  {et~al.}(1996{\natexlab{b}})\citenamefont {Brune}, \citenamefont {Hagley},
  \citenamefont {Dreyer}, \citenamefont {Ma\^{\i}tre}, \citenamefont {Maali},
  \citenamefont {Wunderlich}, \citenamefont {Raimond},\ and\ \citenamefont
  {Haroche}}]{Brune1996}%
  \BibitemOpen
  \bibfield  {author} {\bibinfo {author} {\bibfnamefont {M.}~\bibnamefont
  {Brune}}, \bibinfo {author} {\bibfnamefont {E.}~\bibnamefont {Hagley}},
  \bibinfo {author} {\bibfnamefont {J.}~\bibnamefont {Dreyer}}, \bibinfo
  {author} {\bibfnamefont {X.}~\bibnamefont {Ma\^{\i}tre}}, \bibinfo {author}
  {\bibfnamefont {A.}~\bibnamefont {Maali}}, \bibinfo {author} {\bibfnamefont
  {C.}~\bibnamefont {Wunderlich}}, \bibinfo {author} {\bibfnamefont {J.~M.}\
  \bibnamefont {Raimond}},\ and\ \bibinfo {author} {\bibfnamefont
  {S.}~\bibnamefont {Haroche}},\ }\href
  {https://doi.org/10.1103/PhysRevLett.77.4887} {\bibfield  {journal} {\bibinfo
   {journal} {Phys. Rev. Lett.}\ }\textbf {\bibinfo {volume} {77}},\ \bibinfo
  {pages} {4887} (\bibinfo {year} {1996}{\natexlab{b}})}\BibitemShut {NoStop}%
\bibitem [{\citenamefont {Braunstein}\ and\ \citenamefont
  {Caves}(1994)}]{Braunstein1994}%
  \BibitemOpen
  \bibfield  {author} {\bibinfo {author} {\bibfnamefont {S.~L.}\ \bibnamefont
  {Braunstein}}\ and\ \bibinfo {author} {\bibfnamefont {C.~M.}\ \bibnamefont
  {Caves}},\ }\href {https://doi.org/10.1103/PhysRevLett.72.3439} {\bibfield
  {journal} {\bibinfo  {journal} {Phys. Rev. Lett.}\ }\textbf {\bibinfo
  {volume} {72}},\ \bibinfo {pages} {3439} (\bibinfo {year}
  {1994})}\BibitemShut {NoStop}%
\bibitem [{\citenamefont {Braunstein}\ and\ \citenamefont {van
  Loock}(2005)}]{Braunstein2005}%
  \BibitemOpen
  \bibfield  {author} {\bibinfo {author} {\bibfnamefont {S.~L.}\ \bibnamefont
  {Braunstein}}\ and\ \bibinfo {author} {\bibfnamefont {P.}~\bibnamefont {van
  Loock}},\ }\href {https://doi.org/10.1103/RevModPhys.77.513} {\bibfield
  {journal} {\bibinfo  {journal} {Rev. Mod. Phys.}\ }\textbf {\bibinfo {volume}
  {77}},\ \bibinfo {pages} {513} (\bibinfo {year} {2005})}\BibitemShut
  {NoStop}%
\bibitem [{\citenamefont {Escher}\ \emph {et~al.}(2011)\citenamefont {Escher},
  \citenamefont {de~Matos~Filho},\ and\ \citenamefont
  {Davidovich}}]{Escher2011}%
  \BibitemOpen
  \bibfield  {author} {\bibinfo {author} {\bibfnamefont {B.~M.}\ \bibnamefont
  {Escher}}, \bibinfo {author} {\bibfnamefont {R.~L.}\ \bibnamefont
  {de~Matos~Filho}},\ and\ \bibinfo {author} {\bibfnamefont {L.}~\bibnamefont
  {Davidovich}},\ }\href {https://doi.org/10.1038/nphys1958} {\bibfield
  {journal} {\bibinfo  {journal} {Nat. Phys.}\ }\textbf {\bibinfo {volume}
  {7}},\ \bibinfo {pages} {406} (\bibinfo {year} {2011})}\BibitemShut {NoStop}%
\bibitem [{\citenamefont {Demkowicz-Dobrza{\'n}ski}\ \emph
  {et~al.}(2012)\citenamefont {Demkowicz-Dobrza{\'n}ski}, \citenamefont
  {Ko{\l}ody{\'n}ski},\ and\ \citenamefont {Gu{\c{t}}{\u{a}}}}]{Demkowicz2012}%
  \BibitemOpen
  \bibfield  {author} {\bibinfo {author} {\bibfnamefont {R.}~\bibnamefont
  {Demkowicz-Dobrza{\'n}ski}}, \bibinfo {author} {\bibfnamefont
  {J.}~\bibnamefont {Ko{\l}ody{\'n}ski}},\ and\ \bibinfo {author}
  {\bibfnamefont {M.}~\bibnamefont {Gu{\c{t}}{\u{a}}}},\ }\href
  {https://doi.org/10.1038/ncomms2067} {\bibfield  {journal} {\bibinfo
  {journal} {Nat. Commun.}\ }\textbf {\bibinfo {volume} {3}},\ \bibinfo {pages}
  {1063} (\bibinfo {year} {2012})}\BibitemShut {NoStop}%
\bibitem [{\citenamefont {Zhang}\ and\ \citenamefont {Wu}(2021)}]{ZhangWu2021}%
  \BibitemOpen
  \bibfield  {author} {\bibinfo {author} {\bibfnamefont {Z.-Z.}\ \bibnamefont
  {Zhang}}\ and\ \bibinfo {author} {\bibfnamefont {W.}~\bibnamefont {Wu}},\
  }\href {https://doi.org/10.1103/PhysRevResearch.3.043039} {\bibfield
  {journal} {\bibinfo  {journal} {Phys. Rev. Research}\ }\textbf {\bibinfo
  {volume} {3}},\ \bibinfo {pages} {043039} (\bibinfo {year}
  {2021})}\BibitemShut {NoStop}%
\bibitem [{\citenamefont {Averbukh}\ and\ \citenamefont
  {Perelman}(1989)}]{Averbukh1989}%
  \BibitemOpen
  \bibfield  {author} {\bibinfo {author} {\bibfnamefont {I.}~\bibnamefont
  {Averbukh}}\ and\ \bibinfo {author} {\bibfnamefont {N.}~\bibnamefont
  {Perelman}},\ }\href
  {https://doi.org/https://doi.org/10.1016/0375-9601(89)90943-2} {\bibfield
  {journal} {\bibinfo  {journal} {Physics Letters A}\ }\textbf {\bibinfo
  {volume} {139}},\ \bibinfo {pages} {449} (\bibinfo {year}
  {1989})}\BibitemShut {NoStop}%
\bibitem [{\citenamefont {Averbukh}(1992)}]{Averbukh1992}%
  \BibitemOpen
  \bibfield  {author} {\bibinfo {author} {\bibfnamefont {I.~S.}\ \bibnamefont
  {Averbukh}},\ }\href {https://doi.org/10.1103/PhysRevA.46.R2205} {\bibfield
  {journal} {\bibinfo  {journal} {Phys. Rev. A}\ }\textbf {\bibinfo {volume}
  {46}},\ \bibinfo {pages} {R2205(R)} (\bibinfo {year} {1992})}\BibitemShut
  {NoStop}%
\bibitem [{\citenamefont {Leichtle}\ \emph {et~al.}(1996)\citenamefont
  {Leichtle}, \citenamefont {Averbukh},\ and\ \citenamefont
  {Schleich}}]{Leichtle1996}%
  \BibitemOpen
  \bibfield  {author} {\bibinfo {author} {\bibfnamefont {C.}~\bibnamefont
  {Leichtle}}, \bibinfo {author} {\bibfnamefont {I.~S.}\ \bibnamefont
  {Averbukh}},\ and\ \bibinfo {author} {\bibfnamefont {W.~P.}\ \bibnamefont
  {Schleich}},\ }\href {https://doi.org/10.1103/PhysRevLett.77.3999} {\bibfield
   {journal} {\bibinfo  {journal} {Phys. Rev. Lett.}\ }\textbf {\bibinfo
  {volume} {77}},\ \bibinfo {pages} {3999} (\bibinfo {year}
  {1996})}\BibitemShut {NoStop}%
\bibitem [{\citenamefont {Vrakking}\ \emph {et~al.}(1996)\citenamefont
  {Vrakking}, \citenamefont {Villeneuve},\ and\ \citenamefont
  {Stolow}}]{Vrakking1996}%
  \BibitemOpen
  \bibfield  {author} {\bibinfo {author} {\bibfnamefont {M.~J.~J.}\
  \bibnamefont {Vrakking}}, \bibinfo {author} {\bibfnamefont {D.~M.}\
  \bibnamefont {Villeneuve}},\ and\ \bibinfo {author} {\bibfnamefont
  {A.}~\bibnamefont {Stolow}},\ }\href
  {https://doi.org/10.1103/PhysRevA.54.R37} {\bibfield  {journal} {\bibinfo
  {journal} {Phys. Rev. A}\ }\textbf {\bibinfo {volume} {54}},\ \bibinfo
  {pages} {R37(R)} (\bibinfo {year} {1996})}\BibitemShut {NoStop}%
\bibitem [{\citenamefont {Krantz}\ \emph {et~al.}(2019)\citenamefont {Krantz},
  \citenamefont {Kjaergaard}, \citenamefont {Yan}, \citenamefont {Orlando},
  \citenamefont {Gustavsson},\ and\ \citenamefont {Oliver}}]{Krantz2019}%
  \BibitemOpen
  \bibfield  {author} {\bibinfo {author} {\bibfnamefont {P.}~\bibnamefont
  {Krantz}}, \bibinfo {author} {\bibfnamefont {M.}~\bibnamefont {Kjaergaard}},
  \bibinfo {author} {\bibfnamefont {F.}~\bibnamefont {Yan}}, \bibinfo {author}
  {\bibfnamefont {T.~P.}\ \bibnamefont {Orlando}}, \bibinfo {author}
  {\bibfnamefont {S.}~\bibnamefont {Gustavsson}},\ and\ \bibinfo {author}
  {\bibfnamefont {W.~D.}\ \bibnamefont {Oliver}},\ }\href
  {https://doi.org/10.1063/1.5089550} {\bibfield  {journal} {\bibinfo
  {journal} {Appl. Phys. Rev.}\ }\textbf {\bibinfo {volume} {6}},\ \bibinfo
  {pages} {021318} (\bibinfo {year} {2019})}\BibitemShut {NoStop}%
\bibitem [{\citenamefont {Yan}\ \emph {et~al.}(2018)\citenamefont {Yan},
  \citenamefont {Krantz}, \citenamefont {Sung}, \citenamefont {Kjaergaard},
  \citenamefont {Campbell}, \citenamefont {Orlando}, \citenamefont
  {Gustavsson},\ and\ \citenamefont {Oliver}}]{Yan2018}%
  \BibitemOpen
  \bibfield  {author} {\bibinfo {author} {\bibfnamefont {F.}~\bibnamefont
  {Yan}}, \bibinfo {author} {\bibfnamefont {P.}~\bibnamefont {Krantz}},
  \bibinfo {author} {\bibfnamefont {Y.}~\bibnamefont {Sung}}, \bibinfo {author}
  {\bibfnamefont {M.}~\bibnamefont {Kjaergaard}}, \bibinfo {author}
  {\bibfnamefont {D.~L.}\ \bibnamefont {Campbell}}, \bibinfo {author}
  {\bibfnamefont {T.~P.}\ \bibnamefont {Orlando}}, \bibinfo {author}
  {\bibfnamefont {S.}~\bibnamefont {Gustavsson}},\ and\ \bibinfo {author}
  {\bibfnamefont {W.~D.}\ \bibnamefont {Oliver}},\ }\href
  {https://doi.org/10.1103/PhysRevApplied.10.054062} {\bibfield  {journal}
  {\bibinfo  {journal} {Phys. Rev. Applied}\ }\textbf {\bibinfo {volume}
  {10}},\ \bibinfo {pages} {054062} (\bibinfo {year} {2018})}\BibitemShut
  {NoStop}%
\bibitem [{\citenamefont {Mundada}\ \emph {et~al.}(2019)\citenamefont
  {Mundada}, \citenamefont {Zhang}, \citenamefont {Hazard},\ and\ \citenamefont
  {Houck}}]{Mundada2019}%
  \BibitemOpen
  \bibfield  {author} {\bibinfo {author} {\bibfnamefont {P.}~\bibnamefont
  {Mundada}}, \bibinfo {author} {\bibfnamefont {G.}~\bibnamefont {Zhang}},
  \bibinfo {author} {\bibfnamefont {T.}~\bibnamefont {Hazard}},\ and\ \bibinfo
  {author} {\bibfnamefont {A.}~\bibnamefont {Houck}},\ }\href
  {https://doi.org/10.1103/PhysRevApplied.12.054023} {\bibfield  {journal}
  {\bibinfo  {journal} {Phys. Rev. Applied}\ }\textbf {\bibinfo {volume}
  {12}},\ \bibinfo {pages} {054023} (\bibinfo {year} {2019})}\BibitemShut
  {NoStop}%
\bibitem [{\citenamefont {Lu}\ \emph {et~al.}(2023)\citenamefont {Lu},
  \citenamefont {Kudra}, \citenamefont {Hillmann}, \citenamefont {Yang},
  \citenamefont {Li}, \citenamefont {Quijandr{\'i}a},\ and\ \citenamefont
  {Delsing}}]{Lu2023}%
  \BibitemOpen
  \bibfield  {author} {\bibinfo {author} {\bibfnamefont {Y.}~\bibnamefont
  {Lu}}, \bibinfo {author} {\bibfnamefont {M.}~\bibnamefont {Kudra}}, \bibinfo
  {author} {\bibfnamefont {T.}~\bibnamefont {Hillmann}}, \bibinfo {author}
  {\bibfnamefont {J.}~\bibnamefont {Yang}}, \bibinfo {author} {\bibfnamefont
  {H.-X.}\ \bibnamefont {Li}}, \bibinfo {author} {\bibfnamefont
  {F.}~\bibnamefont {Quijandr{\'i}a}},\ and\ \bibinfo {author} {\bibfnamefont
  {P.}~\bibnamefont {Delsing}},\ }\href
  {https://doi.org/10.1038/s41534-023-00782-w} {\bibfield  {journal} {\bibinfo
  {journal} {npj Quantum Inf.}\ }\textbf {\bibinfo {volume} {9}},\ \bibinfo
  {pages} {114} (\bibinfo {year} {2023})}\BibitemShut {NoStop}%
\bibitem [{\citenamefont {Chapman}\ \emph {et~al.}(2023)\citenamefont
  {Chapman}, \citenamefont {de~Graaf}, \citenamefont {Xue}, \citenamefont
  {Zhang}, \citenamefont {Teoh}, \citenamefont {Curtis}, \citenamefont
  {Tsunoda}, \citenamefont {Eickbusch}, \citenamefont {Read}, \citenamefont
  {Koottandavida}, \citenamefont {Mundhada}, \citenamefont {Frunzio},
  \citenamefont {Devoret}, \citenamefont {Girvin},\ and\ \citenamefont
  {Schoelkopf}}]{Chapman2023}%
  \BibitemOpen
  \bibfield  {author} {\bibinfo {author} {\bibfnamefont {B.~J.}\ \bibnamefont
  {Chapman}}, \bibinfo {author} {\bibfnamefont {S.~J.}\ \bibnamefont
  {de~Graaf}}, \bibinfo {author} {\bibfnamefont {S.~H.}\ \bibnamefont {Xue}},
  \bibinfo {author} {\bibfnamefont {Y.}~\bibnamefont {Zhang}}, \bibinfo
  {author} {\bibfnamefont {J.}~\bibnamefont {Teoh}}, \bibinfo {author}
  {\bibfnamefont {J.~C.}\ \bibnamefont {Curtis}}, \bibinfo {author}
  {\bibfnamefont {T.}~\bibnamefont {Tsunoda}}, \bibinfo {author} {\bibfnamefont
  {A.}~\bibnamefont {Eickbusch}}, \bibinfo {author} {\bibfnamefont {A.~P.}\
  \bibnamefont {Read}}, \bibinfo {author} {\bibfnamefont {A.}~\bibnamefont
  {Koottandavida}}, \bibinfo {author} {\bibfnamefont {S.~O.}\ \bibnamefont
  {Mundhada}}, \bibinfo {author} {\bibfnamefont {L.}~\bibnamefont {Frunzio}},
  \bibinfo {author} {\bibfnamefont {M.~H.}\ \bibnamefont {Devoret}}, \bibinfo
  {author} {\bibfnamefont {S.~M.}\ \bibnamefont {Girvin}},\ and\ \bibinfo
  {author} {\bibfnamefont {R.~J.}\ \bibnamefont {Schoelkopf}},\ }\href
  {https://doi.org/10.1103/PRXQuantum.4.020355} {\bibfield  {journal} {\bibinfo
   {journal} {PRX Quantum}\ }\textbf {\bibinfo {volume} {4}},\ \bibinfo {pages}
  {020355} (\bibinfo {year} {2023})}\BibitemShut {NoStop}%
\bibitem [{\citenamefont {Agarwal}\ and\ \citenamefont
  {Puri}(1989)}]{Agarwal1989}%
  \BibitemOpen
  \bibfield  {author} {\bibinfo {author} {\bibfnamefont {G.~S.}\ \bibnamefont
  {Agarwal}}\ and\ \bibinfo {author} {\bibfnamefont {R.~R.}\ \bibnamefont
  {Puri}},\ }\href {https://doi.org/10.1103/PhysRevA.39.2969} {\bibfield
  {journal} {\bibinfo  {journal} {Phys. Rev. A}\ }\textbf {\bibinfo {volume}
  {39}},\ \bibinfo {pages} {2969} (\bibinfo {year} {1989})}\BibitemShut
  {NoStop}%
\bibitem [{\citenamefont {Bu\v{z}ek}\ and\ \citenamefont
  {Jex}(1990)}]{Buzek1990}%
  \BibitemOpen
  \bibfield  {author} {\bibinfo {author} {\bibfnamefont {V.}~\bibnamefont
  {Bu\v{z}ek}}\ and\ \bibinfo {author} {\bibfnamefont {I.}~\bibnamefont
  {Jex}},\ }\href {https://doi.org/10.1016/0030-4018(90)90340-Y} {\bibfield
  {journal} {\bibinfo  {journal} {Opt. Commun.}\ }\textbf {\bibinfo {volume}
  {78}},\ \bibinfo {pages} {425} (\bibinfo {year} {1990})}\BibitemShut
  {NoStop}%
\bibitem [{\citenamefont {Joshi}\ and\ \citenamefont
  {Puri}(1992)}]{Joshi1992a}%
  \BibitemOpen
  \bibfield  {author} {\bibinfo {author} {\bibfnamefont {A.}~\bibnamefont
  {Joshi}}\ and\ \bibinfo {author} {\bibfnamefont {R.~R.}\ \bibnamefont
  {Puri}},\ }\href {https://doi.org/10.1103/PhysRevA.45.5056} {\bibfield
  {journal} {\bibinfo  {journal} {Phys. Rev. A}\ }\textbf {\bibinfo {volume}
  {45}},\ \bibinfo {pages} {5056} (\bibinfo {year} {1992})}\BibitemShut
  {NoStop}%
\bibitem [{\citenamefont {Werner}\ and\ \citenamefont
  {Risken}(1991)}]{Werner1991}%
  \BibitemOpen
  \bibfield  {author} {\bibinfo {author} {\bibfnamefont {M.~J.}\ \bibnamefont
  {Werner}}\ and\ \bibinfo {author} {\bibfnamefont {H.}~\bibnamefont
  {Risken}},\ }\href {https://doi.org/10.1103/PhysRevA.44.4623} {\bibfield
  {journal} {\bibinfo  {journal} {Phys. Rev. A}\ }\textbf {\bibinfo {volume}
  {44}},\ \bibinfo {pages} {4623} (\bibinfo {year} {1991})}\BibitemShut
  {NoStop}%
\bibitem [{\citenamefont {Baghshahi}\ \emph {et~al.}(2014)\citenamefont
  {Baghshahi}, \citenamefont {Tavassoly},\ and\ \citenamefont
  {Behjat}}]{Baghshahi2014}%
  \BibitemOpen
  \bibfield  {author} {\bibinfo {author} {\bibfnamefont {H.~R.}\ \bibnamefont
  {Baghshahi}}, \bibinfo {author} {\bibfnamefont {M.~K.}\ \bibnamefont
  {Tavassoly}},\ and\ \bibinfo {author} {\bibfnamefont {A.}~\bibnamefont
  {Behjat}},\ }\href {https://doi.org/10.1088/1674-1056/23/7/074203} {\bibfield
   {journal} {\bibinfo  {journal} {Chin. Phys. B}\ }\textbf {\bibinfo {volume}
  {23}},\ \bibinfo {pages} {074203} (\bibinfo {year} {2014})}\BibitemShut
  {NoStop}%
\bibitem [{\citenamefont {Ashefas}\ \emph {et~al.}(2022)\citenamefont
  {Ashefas}, \citenamefont {Manosh},\ and\ \citenamefont
  {Thayyullathil}}]{Ashefas2022}%
  \BibitemOpen
  \bibfield  {author} {\bibinfo {author} {\bibfnamefont {C.~H.~M.}\
  \bibnamefont {Ashefas}}, \bibinfo {author} {\bibfnamefont {T.~M.}\
  \bibnamefont {Manosh}},\ and\ \bibinfo {author} {\bibfnamefont {R.~B.}\
  \bibnamefont {Thayyullathil}},\ }\href
  {https://doi.org/10.1007/s10773-022-05173-z} {\bibfield  {journal} {\bibinfo
  {journal} {Int. J. Theor. Phys.}\ }\textbf {\bibinfo {volume} {61}},\
  \bibinfo {pages} {186} (\bibinfo {year} {2022})}\BibitemShut {NoStop}%
\bibitem [{\citenamefont {Bandyopadhyay}\ and\ \citenamefont
  {Gangopadhyay}(1996)}]{Bandyopadhyay1996}%
  \BibitemOpen
  \bibfield  {author} {\bibinfo {author} {\bibfnamefont {A.}~\bibnamefont
  {Bandyopadhyay}}\ and\ \bibinfo {author} {\bibfnamefont {G.}~\bibnamefont
  {Gangopadhyay}},\ }\href {https://doi.org/10.1080/09500349608232759}
  {\bibfield  {journal} {\bibinfo  {journal} {J. Mod. Opt.}\ }\textbf {\bibinfo
  {volume} {43}},\ \bibinfo {pages} {487} (\bibinfo {year} {1996})}\BibitemShut
  {NoStop}%
\bibitem [{\citenamefont {Bocanegra-Garay}\ \emph {et~al.}(2024)\citenamefont
  {Bocanegra-Garay}, \citenamefont {Castillo-Celeita}, \citenamefont {Negro},
  \citenamefont {Nieto},\ and\ \citenamefont {G\'omez-Ruiz}}]{Bocanegra2024}%
  \BibitemOpen
  \bibfield  {author} {\bibinfo {author} {\bibfnamefont {I.~A.}\ \bibnamefont
  {Bocanegra-Garay}}, \bibinfo {author} {\bibfnamefont {M.}~\bibnamefont
  {Castillo-Celeita}}, \bibinfo {author} {\bibfnamefont {J.}~\bibnamefont
  {Negro}}, \bibinfo {author} {\bibfnamefont {L.~M.}\ \bibnamefont {Nieto}},\
  and\ \bibinfo {author} {\bibfnamefont {F.~J.}\ \bibnamefont {G\'omez-Ruiz}},\
  }\href {https://doi.org/10.1103/PhysRevResearch.6.043218} {\bibfield
  {journal} {\bibinfo  {journal} {Phys. Rev. Res.}\ }\textbf {\bibinfo {volume}
  {6}},\ \bibinfo {pages} {043218} (\bibinfo {year} {2024})}\BibitemShut
  {NoStop}%
\bibitem [{\citenamefont {Bocanegra-Garay}\ \emph {et~al.}(2025)\citenamefont
  {Bocanegra-Garay}, \citenamefont {Sánchez},\ and\ \citenamefont
  {Moya-Cessa}}]{Bocanegra2025}%
  \BibitemOpen
  \bibfield  {author} {\bibinfo {author} {\bibfnamefont {I.~A.}\ \bibnamefont
  {Bocanegra-Garay}}, \bibinfo {author} {\bibfnamefont {L.~H.}\ \bibnamefont
  {Sánchez}},\ and\ \bibinfo {author} {\bibfnamefont {H.~M.}\ \bibnamefont
  {Moya-Cessa}},\ }\href {https://arxiv.org/abs/2501.14907} {\  (\bibinfo
  {year} {2025})},\ \Eprint {https://arxiv.org/abs/2501.14907}
  {arXiv:2501.14907 [quant-ph]} \BibitemShut {NoStop}%
\bibitem [{\citenamefont {Burgarth}\ \emph {et~al.}(2024)\citenamefont
  {Burgarth}, \citenamefont {de~Beaudrap}, \citenamefont {Pallister},\ and\
  \citenamefont {Regula}}]{Burgarth2024}%
  \BibitemOpen
  \bibfield  {author} {\bibinfo {author} {\bibfnamefont {D.}~\bibnamefont
  {Burgarth}}, \bibinfo {author} {\bibfnamefont {N.}~\bibnamefont
  {de~Beaudrap}}, \bibinfo {author} {\bibfnamefont {S.}~\bibnamefont
  {Pallister}},\ and\ \bibinfo {author} {\bibfnamefont {B.}~\bibnamefont
  {Regula}},\ }\href {https://doi.org/10.1103/PhysRevA.109.013702} {\bibfield
  {journal} {\bibinfo  {journal} {Phys. Rev. A}\ }\textbf {\bibinfo {volume}
  {109}},\ \bibinfo {pages} {013702} (\bibinfo {year} {2024})}\BibitemShut
  {NoStop}%
\bibitem [{\citenamefont {Mangazeev}\ \emph {et~al.}(2021)\citenamefont
  {Mangazeev}, \citenamefont {Batchelor},\ and\ \citenamefont
  {Bazhanov}}]{Mangazeev2021}%
  \BibitemOpen
  \bibfield  {author} {\bibinfo {author} {\bibfnamefont {V.~V.}\ \bibnamefont
  {Mangazeev}}, \bibinfo {author} {\bibfnamefont {M.~T.}\ \bibnamefont
  {Batchelor}},\ and\ \bibinfo {author} {\bibfnamefont {V.~V.}\ \bibnamefont
  {Bazhanov}},\ }\href {https://doi.org/10.1088/1751-8121/abe426} {\bibfield
  {journal} {\bibinfo  {journal} {Journal of Physics A: Mathematical and
  Theoretical}\ }\textbf {\bibinfo {volume} {54}},\ \bibinfo {pages} {12LT01}
  (\bibinfo {year} {2021})}\BibitemShut {NoStop}%
\bibitem [{\citenamefont {Reyes-Bustos}\ \emph {et~al.}(2021)\citenamefont
  {Reyes-Bustos}, \citenamefont {Braak},\ and\ \citenamefont
  {Wakayama}}]{Reyes2021}%
  \BibitemOpen
  \bibfield  {author} {\bibinfo {author} {\bibfnamefont {C.}~\bibnamefont
  {Reyes-Bustos}}, \bibinfo {author} {\bibfnamefont {D.}~\bibnamefont
  {Braak}},\ and\ \bibinfo {author} {\bibfnamefont {M.}~\bibnamefont
  {Wakayama}},\ }\href {https://doi.org/10.1088/1751-8121/ac0508} {\bibfield
  {journal} {\bibinfo  {journal} {Journal of Physics A: Mathematical and
  Theoretical}\ }\textbf {\bibinfo {volume} {54}},\ \bibinfo {pages} {285202}
  (\bibinfo {year} {2021})}\BibitemShut {NoStop}%
\bibitem [{\citenamefont {Omolo}(2022{\natexlab{b}})}]{Omolo2022b}%
  \BibitemOpen
  \bibfield  {author} {\bibinfo {author} {\bibfnamefont {J.~A.}\ \bibnamefont
  {Omolo}},\ }\href {https://arxiv.org/abs/2207.02730} {\  (\bibinfo {year}
  {2022}{\natexlab{b}})},\ \Eprint {https://arxiv.org/abs/2207.02730}
  {arXiv:2207.02730 [quant-ph]} \BibitemShut {NoStop}%
\bibitem [{\citenamefont {Bina}\ \emph {et~al.}(2014)\citenamefont {Bina},
  \citenamefont {Amelio},\ and\ \citenamefont {Paris}}]{Bina2014}%
  \BibitemOpen
  \bibfield  {author} {\bibinfo {author} {\bibfnamefont {M.}~\bibnamefont
  {Bina}}, \bibinfo {author} {\bibfnamefont {I.}~\bibnamefont {Amelio}},\ and\
  \bibinfo {author} {\bibfnamefont {M.~G.~A.}\ \bibnamefont {Paris}},\ }\href
  {https://doi.org/10.1103/PhysRevA.89.042118} {\bibfield  {journal} {\bibinfo
  {journal} {Phys. Rev. A}\ }\textbf {\bibinfo {volume} {89}},\ \bibinfo
  {pages} {042118} (\bibinfo {year} {2014})}\BibitemShut {NoStop}%
\bibitem [{\citenamefont {Birnbaum}\ \emph {et~al.}(2005)\citenamefont
  {Birnbaum}, \citenamefont {Boca}, \citenamefont {Miller}, \citenamefont
  {Boozer}, \citenamefont {Northup},\ and\ \citenamefont
  {Kimble}}]{Birnbaum2005}%
  \BibitemOpen
  \bibfield  {author} {\bibinfo {author} {\bibfnamefont {K.~M.}\ \bibnamefont
  {Birnbaum}}, \bibinfo {author} {\bibfnamefont {A.}~\bibnamefont {Boca}},
  \bibinfo {author} {\bibfnamefont {R.}~\bibnamefont {Miller}}, \bibinfo
  {author} {\bibfnamefont {A.~D.}\ \bibnamefont {Boozer}}, \bibinfo {author}
  {\bibfnamefont {T.~E.}\ \bibnamefont {Northup}},\ and\ \bibinfo {author}
  {\bibfnamefont {H.~J.}\ \bibnamefont {Kimble}},\ }\href
  {https://doi.org/10.1038/nature03804} {\bibfield  {journal} {\bibinfo
  {journal} {Nature}\ }\textbf {\bibinfo {volume} {436}},\ \bibinfo {pages}
  {87} (\bibinfo {year} {2005})}\BibitemShut {NoStop}%
\bibitem [{\citenamefont {Dong}\ \emph {et~al.}(2026)\citenamefont {Dong},
  \citenamefont {Shah}, \citenamefont {Kirton}, \citenamefont {Alaeian},\ and\
  \citenamefont {Felicetti}}]{Lijuan2026}%
  \BibitemOpen
  \bibfield  {author} {\bibinfo {author} {\bibfnamefont {L.}~\bibnamefont
  {Dong}}, \bibinfo {author} {\bibfnamefont {A.~J.}\ \bibnamefont {Shah}},
  \bibinfo {author} {\bibfnamefont {P.}~\bibnamefont {Kirton}}, \bibinfo
  {author} {\bibfnamefont {H.}~\bibnamefont {Alaeian}},\ and\ \bibinfo {author}
  {\bibfnamefont {S.}~\bibnamefont {Felicetti}},\ }\href
  {https://doi.org/10.1103/l477-z8jr} {\bibfield  {journal} {\bibinfo
  {journal} {PRX Quantum}\ }\textbf {\bibinfo {volume} {7}},\ \bibinfo {pages}
  {020349} (\bibinfo {year} {2026})}\BibitemShut {NoStop}%
\bibitem [{\citenamefont {Ng}\ \emph {et~al.}(1999)\citenamefont {Ng},
  \citenamefont {Lo},\ and\ \citenamefont {Liu}}]{Ng1999}%
  \BibitemOpen
  \bibfield  {author} {\bibinfo {author} {\bibfnamefont {K.~M.}\ \bibnamefont
  {Ng}}, \bibinfo {author} {\bibfnamefont {C.~F.}\ \bibnamefont {Lo}},\ and\
  \bibinfo {author} {\bibfnamefont {K.~L.}\ \bibnamefont {Liu}},\ }\href
  {https://doi.org/10.1007/PL00021611} {\bibfield  {journal} {\bibinfo
  {journal} {Eur. Phys. J. D}\ }\textbf {\bibinfo {volume} {6}},\ \bibinfo
  {pages} {119} (\bibinfo {year} {1999})}\BibitemShut {NoStop}%
\bibitem [{\citenamefont {Zhang}\ and\ \citenamefont {Chen}(2015)}]{Zhang2015}%
  \BibitemOpen
  \bibfield  {author} {\bibinfo {author} {\bibfnamefont {Y.-Y.}\ \bibnamefont
  {Zhang}}\ and\ \bibinfo {author} {\bibfnamefont {Q.-H.}\ \bibnamefont
  {Chen}},\ }\href {https://doi.org/10.1103/PhysRevA.91.013814} {\bibfield
  {journal} {\bibinfo  {journal} {Phys. Rev. A}\ }\textbf {\bibinfo {volume}
  {91}},\ \bibinfo {pages} {013814} (\bibinfo {year} {2015})}\BibitemShut
  {NoStop}%
\bibitem [{\citenamefont {Sen}\ \emph {et~al.}(2019)\citenamefont {Sen},
  \citenamefont {Dey},\ and\ \citenamefont {Nath}}]{Sen2019}%
  \BibitemOpen
  \bibfield  {author} {\bibinfo {author} {\bibfnamefont {S.}~\bibnamefont
  {Sen}}, \bibinfo {author} {\bibfnamefont {T.~K.}\ \bibnamefont {Dey}},\ and\
  \bibinfo {author} {\bibfnamefont {M.~R.}\ \bibnamefont {Nath}},\ }\href
  {https://doi.org/10.1140/epjd/e2018-90372-7} {\bibfield  {journal} {\bibinfo
  {journal} {Eur. Phys. J. D}\ }\textbf {\bibinfo {volume} {73}},\ \bibinfo
  {pages} {3} (\bibinfo {year} {2019})}\BibitemShut {NoStop}%
\bibitem [{\citenamefont {Yu}\ and\ \citenamefont {Eberly}(2004)}]{Yu2004}%
  \BibitemOpen
  \bibfield  {author} {\bibinfo {author} {\bibfnamefont {T.}~\bibnamefont
  {Yu}}\ and\ \bibinfo {author} {\bibfnamefont {J.~H.}\ \bibnamefont
  {Eberly}},\ }\href {https://doi.org/10.1103/PhysRevLett.93.140404} {\bibfield
   {journal} {\bibinfo  {journal} {Phys. Rev. Lett.}\ }\textbf {\bibinfo
  {volume} {93}},\ \bibinfo {pages} {140404} (\bibinfo {year}
  {2004})}\BibitemShut {NoStop}%
\bibitem [{\citenamefont {Clerk}\ \emph {et~al.}(2010)\citenamefont {Clerk},
  \citenamefont {Devoret}, \citenamefont {Girvin}, \citenamefont {Marquardt},\
  and\ \citenamefont {Schoelkopf}}]{Clerk2010}%
  \BibitemOpen
  \bibfield  {author} {\bibinfo {author} {\bibfnamefont {A.~A.}\ \bibnamefont
  {Clerk}}, \bibinfo {author} {\bibfnamefont {M.~H.}\ \bibnamefont {Devoret}},
  \bibinfo {author} {\bibfnamefont {S.~M.}\ \bibnamefont {Girvin}}, \bibinfo
  {author} {\bibfnamefont {F.}~\bibnamefont {Marquardt}},\ and\ \bibinfo
  {author} {\bibfnamefont {R.~J.}\ \bibnamefont {Schoelkopf}},\ }\href
  {https://doi.org/10.1103/RevModPhys.82.1155} {\bibfield  {journal} {\bibinfo
  {journal} {Rev. Mod. Phys.}\ }\textbf {\bibinfo {volume} {82}},\ \bibinfo
  {pages} {1155} (\bibinfo {year} {2010})}\BibitemShut {NoStop}%
\bibitem [{\citenamefont {Gerritsma}\ \emph {et~al.}(2010)\citenamefont
  {Gerritsma}, \citenamefont {Kirchmair}, \citenamefont {Z\"ahringer},
  \citenamefont {Solano}, \citenamefont {Blatt},\ and\ \citenamefont
  {Roos}}]{Gerritsma2010}%
  \BibitemOpen
  \bibfield  {author} {\bibinfo {author} {\bibfnamefont {R.}~\bibnamefont
  {Gerritsma}}, \bibinfo {author} {\bibfnamefont {G.}~\bibnamefont
  {Kirchmair}}, \bibinfo {author} {\bibfnamefont {F.}~\bibnamefont
  {Z\"ahringer}}, \bibinfo {author} {\bibfnamefont {E.}~\bibnamefont {Solano}},
  \bibinfo {author} {\bibfnamefont {R.}~\bibnamefont {Blatt}},\ and\ \bibinfo
  {author} {\bibfnamefont {C.~F.}\ \bibnamefont {Roos}},\ }\href
  {https://doi.org/10.1038/nature08688} {\bibfield  {journal} {\bibinfo
  {journal} {Nature}\ }\textbf {\bibinfo {volume} {463}},\ \bibinfo {pages}
  {68} (\bibinfo {year} {2010})}\BibitemShut {NoStop}%
\bibitem [{\citenamefont {Yan}\ \emph {et~al.}(2013)\citenamefont {Yan},
  \citenamefont {Gustavsson}, \citenamefont {Bylander}, \citenamefont {Jin},
  \citenamefont {Yoshihara}, \citenamefont {Cory}, \citenamefont {Nakamura},
  \citenamefont {Orlando},\ and\ \citenamefont {Oliver}}]{Yan2013}%
  \BibitemOpen
  \bibfield  {author} {\bibinfo {author} {\bibfnamefont {F.}~\bibnamefont
  {Yan}}, \bibinfo {author} {\bibfnamefont {S.}~\bibnamefont {Gustavsson}},
  \bibinfo {author} {\bibfnamefont {J.}~\bibnamefont {Bylander}}, \bibinfo
  {author} {\bibfnamefont {X.}~\bibnamefont {Jin}}, \bibinfo {author}
  {\bibfnamefont {F.}~\bibnamefont {Yoshihara}}, \bibinfo {author}
  {\bibfnamefont {D.~G.}\ \bibnamefont {Cory}}, \bibinfo {author}
  {\bibfnamefont {Y.}~\bibnamefont {Nakamura}}, \bibinfo {author}
  {\bibfnamefont {T.~P.}\ \bibnamefont {Orlando}},\ and\ \bibinfo {author}
  {\bibfnamefont {W.~D.}\ \bibnamefont {Oliver}},\ }\href
  {https://doi.org/10.1038/ncomms3337} {\bibfield  {journal} {\bibinfo
  {journal} {Nature Communications}\ }\textbf {\bibinfo {volume} {4}},\
  \bibinfo {pages} {2337} (\bibinfo {year} {2013})}\BibitemShut {NoStop}%
\bibitem [{\citenamefont {Alvermann}\ \emph {et~al.}(2012)\citenamefont
  {Alvermann}, \citenamefont {Bakemeier},\ and\ \citenamefont
  {Fehske}}]{Alvermann2012}%
  \BibitemOpen
  \bibfield  {author} {\bibinfo {author} {\bibfnamefont {A.}~\bibnamefont
  {Alvermann}}, \bibinfo {author} {\bibfnamefont {L.}~\bibnamefont
  {Bakemeier}},\ and\ \bibinfo {author} {\bibfnamefont {H.}~\bibnamefont
  {Fehske}},\ }\href {https://doi.org/10.1103/PhysRevA.85.043803} {\bibfield
  {journal} {\bibinfo  {journal} {Phys. Rev. A}\ }\textbf {\bibinfo {volume}
  {85}},\ \bibinfo {pages} {043803} (\bibinfo {year} {2012})}\BibitemShut
  {NoStop}%
\bibitem [{\citenamefont {Anaya~Contreras}\ \emph {et~al.}(2025)\citenamefont
  {Anaya~Contreras}, \citenamefont {Ramos~Prieto}, \citenamefont
  {Zuniga~Segundo},\ and\ \citenamefont {Moya~Cessa}}]{Anaya2025}%
  \BibitemOpen
  \bibfield  {author} {\bibinfo {author} {\bibfnamefont {J.~A.}\ \bibnamefont
  {Anaya~Contreras}}, \bibinfo {author} {\bibfnamefont {I.}~\bibnamefont
  {Ramos~Prieto}}, \bibinfo {author} {\bibfnamefont {A.}~\bibnamefont
  {Zuniga~Segundo}},\ and\ \bibinfo {author} {\bibfnamefont {H.~M.}\
  \bibnamefont {Moya~Cessa}},\ }\bibfield  {journal} {\bibinfo  {journal}
  {Front. Phys.}\ }\textbf {\bibinfo {volume} {13}},\ \href
  {https://doi.org/10.3389/fphy.2025.1568407} {10.3389/fphy.2025.1568407}
  (\bibinfo {year} {2025})\BibitemShut {NoStop}%
\bibitem [{\citenamefont {Ashhab}(2013)}]{Ashhab2013}%
  \BibitemOpen
  \bibfield  {author} {\bibinfo {author} {\bibfnamefont {S.}~\bibnamefont
  {Ashhab}},\ }\href {https://doi.org/10.1103/PhysRevA.87.013826} {\bibfield
  {journal} {\bibinfo  {journal} {Phys. Rev. A}\ }\textbf {\bibinfo {volume}
  {87}},\ \bibinfo {pages} {013826} (\bibinfo {year} {2013})}\BibitemShut
  {NoStop}%
\bibitem [{\citenamefont {Berry}(1984)}]{Berry1984}%
  \BibitemOpen
  \bibfield  {author} {\bibinfo {author} {\bibfnamefont {M.~V.}\ \bibnamefont
  {Berry}},\ }\href {https://doi.org/10.1098/rspa.1984.0023} {\bibfield
  {journal} {\bibinfo  {journal} {Proc. R. Soc. Lond. A}\ }\textbf {\bibinfo
  {volume} {392}},\ \bibinfo {pages} {45} (\bibinfo {year} {1984})}\BibitemShut
  {NoStop}%
\bibitem [{\citenamefont {Aharonov}\ and\ \citenamefont
  {Anandan}(1987)}]{Aharonov1987}%
  \BibitemOpen
  \bibfield  {author} {\bibinfo {author} {\bibfnamefont {Y.}~\bibnamefont
  {Aharonov}}\ and\ \bibinfo {author} {\bibfnamefont {J.}~\bibnamefont
  {Anandan}},\ }\href {https://doi.org/10.1103/PhysRevLett.58.1593} {\bibfield
  {journal} {\bibinfo  {journal} {Phys. Rev. Lett.}\ }\textbf {\bibinfo
  {volume} {58}},\ \bibinfo {pages} {1593} (\bibinfo {year}
  {1987})}\BibitemShut {NoStop}%
\bibitem [{\citenamefont {Bloch}\ and\ \citenamefont
  {Siegert}(1940)}]{Bloch1940}%
  \BibitemOpen
  \bibfield  {author} {\bibinfo {author} {\bibfnamefont {F.}~\bibnamefont
  {Bloch}}\ and\ \bibinfo {author} {\bibfnamefont {A.}~\bibnamefont
  {Siegert}},\ }\href {https://doi.org/10.1103/PhysRev.57.522} {\bibfield
  {journal} {\bibinfo  {journal} {Phys. Rev.}\ }\textbf {\bibinfo {volume}
  {57}},\ \bibinfo {pages} {522} (\bibinfo {year} {1940})}\BibitemShut
  {NoStop}%
\bibitem [{\citenamefont {Ragy}\ \emph {et~al.}(2016)\citenamefont {Ragy},
  \citenamefont {Jarzyna},\ and\ \citenamefont
  {Demkowicz-Dobrza\'nski}}]{Ragy2016}%
  \BibitemOpen
  \bibfield  {author} {\bibinfo {author} {\bibfnamefont {S.}~\bibnamefont
  {Ragy}}, \bibinfo {author} {\bibfnamefont {M.}~\bibnamefont {Jarzyna}},\ and\
  \bibinfo {author} {\bibfnamefont {R.}~\bibnamefont
  {Demkowicz-Dobrza\'nski}},\ }\href
  {https://doi.org/10.1103/PhysRevA.94.052108} {\bibfield  {journal} {\bibinfo
  {journal} {Phys. Rev. A}\ }\textbf {\bibinfo {volume} {94}},\ \bibinfo
  {pages} {052108} (\bibinfo {year} {2016})}\BibitemShut {NoStop}%
\bibitem [{\citenamefont {Kull}\ \emph {et~al.}(2020)\citenamefont {Kull},
  \citenamefont {Gu\'erin},\ and\ \citenamefont {Verstraete}}]{Kull2020}%
  \BibitemOpen
  \bibfield  {author} {\bibinfo {author} {\bibfnamefont {I.}~\bibnamefont
  {Kull}}, \bibinfo {author} {\bibfnamefont {P.~A.}\ \bibnamefont {Gu\'erin}},\
  and\ \bibinfo {author} {\bibfnamefont {F.}~\bibnamefont {Verstraete}},\
  }\href {https://doi.org/10.1088/1751-8121/ab8416} {\bibfield  {journal}
  {\bibinfo  {journal} {J. Phys. A: Math. Theor.}\ }\textbf {\bibinfo {volume}
  {53}},\ \bibinfo {pages} {244001} (\bibinfo {year} {2020})}\BibitemShut
  {NoStop}%
\bibitem [{\citenamefont {Seveso}\ \emph {et~al.}(2017)\citenamefont {Seveso},
  \citenamefont {Rossi},\ and\ \citenamefont {Paris}}]{Seveso2017}%
  \BibitemOpen
  \bibfield  {author} {\bibinfo {author} {\bibfnamefont {L.}~\bibnamefont
  {Seveso}}, \bibinfo {author} {\bibfnamefont {M.~A.~C.}\ \bibnamefont
  {Rossi}},\ and\ \bibinfo {author} {\bibfnamefont {M.~G.~A.}\ \bibnamefont
  {Paris}},\ }\href {https://doi.org/10.1103/PhysRevA.95.012111} {\bibfield
  {journal} {\bibinfo  {journal} {Phys. Rev. A}\ }\textbf {\bibinfo {volume}
  {95}},\ \bibinfo {pages} {012111} (\bibinfo {year} {2017})}\BibitemShut
  {NoStop}%
\bibitem [{\citenamefont {Szczykulska}\ \emph {et~al.}(2016)\citenamefont
  {Szczykulska}, \citenamefont {Baumgratz},\ and\ \citenamefont
  {Datta}}]{Szczykulska2016}%
  \BibitemOpen
  \bibfield  {author} {\bibinfo {author} {\bibfnamefont {M.}~\bibnamefont
  {Szczykulska}}, \bibinfo {author} {\bibfnamefont {T.}~\bibnamefont
  {Baumgratz}},\ and\ \bibinfo {author} {\bibfnamefont {A.}~\bibnamefont
  {Datta}},\ }\href {https://doi.org/10.1080/23746149.2016.1230476} {\bibfield
  {journal} {\bibinfo  {journal} {Advances in Physics: X}\ }\textbf {\bibinfo
  {volume} {1}},\ \bibinfo {pages} {621} (\bibinfo {year} {2016})}\BibitemShut
  {NoStop}%
\bibitem [{\citenamefont {Liu}\ \emph {et~al.}(2020)\citenamefont {Liu},
  \citenamefont {Yuan}, \citenamefont {Lu},\ and\ \citenamefont
  {Wang}}]{Liu2020}%
  \BibitemOpen
  \bibfield  {author} {\bibinfo {author} {\bibfnamefont {J.}~\bibnamefont
  {Liu}}, \bibinfo {author} {\bibfnamefont {H.}~\bibnamefont {Yuan}}, \bibinfo
  {author} {\bibfnamefont {X.-M.}\ \bibnamefont {Lu}},\ and\ \bibinfo {author}
  {\bibfnamefont {X.}~\bibnamefont {Wang}},\ }\href
  {https://doi.org/10.1088/1751-8121/ab5d4d} {\bibfield  {journal} {\bibinfo
  {journal} {J. Phys. A: Math. Theor.}\ }\textbf {\bibinfo {volume} {53}},\
  \bibinfo {pages} {023001} (\bibinfo {year} {2020})}\BibitemShut {NoStop}%
\bibitem [{\citenamefont {Chen}\ \emph {et~al.}(2022)\citenamefont {Chen},
  \citenamefont {Chen},\ and\ \citenamefont {Yuan}}]{Chen2022}%
  \BibitemOpen
  \bibfield  {author} {\bibinfo {author} {\bibfnamefont {H.}~\bibnamefont
  {Chen}}, \bibinfo {author} {\bibfnamefont {Y.}~\bibnamefont {Chen}},\ and\
  \bibinfo {author} {\bibfnamefont {H.}~\bibnamefont {Yuan}},\ }\href
  {https://doi.org/10.1103/PhysRevLett.128.250502} {\bibfield  {journal}
  {\bibinfo  {journal} {Phys. Rev. Lett.}\ }\textbf {\bibinfo {volume} {128}},\
  \bibinfo {pages} {250502} (\bibinfo {year} {2022})}\BibitemShut {NoStop}%
\bibitem [{\citenamefont {He}\ and\ \citenamefont {Paris}(2025)}]{HeParis2025}%
  \BibitemOpen
  \bibfield  {author} {\bibinfo {author} {\bibfnamefont {J.}~\bibnamefont
  {He}}\ and\ \bibinfo {author} {\bibfnamefont {M.~G.~A.}\ \bibnamefont
  {Paris}},\ }\href {https://doi.org/10.1088/1751-8121/adf585} {\bibfield
  {journal} {\bibinfo  {journal} {Journal of Physics A: Mathematical and
  Theoretical}\ }\textbf {\bibinfo {volume} {58}},\ \bibinfo {pages} {325301}
  (\bibinfo {year} {2025})}\BibitemShut {NoStop}%
\bibitem [{\citenamefont {Frigerio}\ and\ \citenamefont
  {Paris}(2026)}]{Frigerio2025}%
  \BibitemOpen
  \bibfield  {author} {\bibinfo {author} {\bibfnamefont {M.}~\bibnamefont
  {Frigerio}}\ and\ \bibinfo {author} {\bibfnamefont {M.~G.~A.}\ \bibnamefont
  {Paris}},\ }\href {https://doi.org/10.1142/S0219749925400015} {\bibfield
  {journal} {\bibinfo  {journal} {International Journal of Quantum
  Information}\ }\textbf {\bibinfo {volume} {24}},\ \bibinfo {pages} {2540001}
  (\bibinfo {year} {2026})}\BibitemShut {NoStop}%
\bibitem [{\citenamefont {Wineland}\ \emph {et~al.}(1992)\citenamefont
  {Wineland}, \citenamefont {Bollinger}, \citenamefont {Itano}, \citenamefont
  {Moore},\ and\ \citenamefont {Heinzen}}]{Wineland1992}%
  \BibitemOpen
  \bibfield  {author} {\bibinfo {author} {\bibfnamefont {D.~J.}\ \bibnamefont
  {Wineland}}, \bibinfo {author} {\bibfnamefont {J.~J.}\ \bibnamefont
  {Bollinger}}, \bibinfo {author} {\bibfnamefont {W.~M.}\ \bibnamefont
  {Itano}}, \bibinfo {author} {\bibfnamefont {F.~L.}\ \bibnamefont {Moore}},\
  and\ \bibinfo {author} {\bibfnamefont {D.~J.}\ \bibnamefont {Heinzen}},\
  }\href {https://doi.org/10.1103/PhysRevA.46.R6797} {\bibfield  {journal}
  {\bibinfo  {journal} {Phys. Rev. A}\ }\textbf {\bibinfo {volume} {46}},\
  \bibinfo {pages} {R6797} (\bibinfo {year} {1992})}\BibitemShut {NoStop}%
\bibitem [{\citenamefont {Holland}\ and\ \citenamefont
  {Burnett}(1993)}]{Holland1993}%
  \BibitemOpen
  \bibfield  {author} {\bibinfo {author} {\bibfnamefont {M.~J.}\ \bibnamefont
  {Holland}}\ and\ \bibinfo {author} {\bibfnamefont {K.}~\bibnamefont
  {Burnett}},\ }\href {https://doi.org/10.1103/PhysRevLett.71.1355} {\bibfield
  {journal} {\bibinfo  {journal} {Phys. Rev. Lett.}\ }\textbf {\bibinfo
  {volume} {71}},\ \bibinfo {pages} {1355} (\bibinfo {year}
  {1993})}\BibitemShut {NoStop}%
\bibitem [{\citenamefont {Gutenkunst}\ \emph {et~al.}(2007)\citenamefont
  {Gutenkunst}, \citenamefont {Waterfall}, \citenamefont {Casey}, \citenamefont
  {Brown}, \citenamefont {Myers},\ and\ \citenamefont
  {Sethna}}]{Gutenkunst2007}%
  \BibitemOpen
  \bibfield  {author} {\bibinfo {author} {\bibfnamefont {R.~N.}\ \bibnamefont
  {Gutenkunst}}, \bibinfo {author} {\bibfnamefont {J.~J.}\ \bibnamefont
  {Waterfall}}, \bibinfo {author} {\bibfnamefont {F.~P.}\ \bibnamefont
  {Casey}}, \bibinfo {author} {\bibfnamefont {K.~S.}\ \bibnamefont {Brown}},
  \bibinfo {author} {\bibfnamefont {C.~R.}\ \bibnamefont {Myers}},\ and\
  \bibinfo {author} {\bibfnamefont {J.~P.}\ \bibnamefont {Sethna}},\ }\href
  {https://doi.org/10.1371/journal.pcbi.0030189} {\bibfield  {journal}
  {\bibinfo  {journal} {PLoS Comput. Biol.}\ }\textbf {\bibinfo {volume} {3}},\
  \bibinfo {pages} {e189} (\bibinfo {year} {2007})}\BibitemShut {NoStop}%
\bibitem [{\citenamefont {Hayashi}(2005)}]{Hayashi2005}%
  \BibitemOpen
  \bibinfo {editor} {\bibfnamefont {M.}~\bibnamefont {Hayashi}},\ ed.,\
  \href@noop {} {\emph {\bibinfo {title} {Asymptotic Theory of Quantum
  Statistical Inference}}}\ (\bibinfo  {publisher} {World Scientific},\
  \bibinfo {address} {Singapore},\ \bibinfo {year} {2005})\BibitemShut
  {NoStop}%
\end{thebibliography}

%

\end{document}